\documentclass[a4paper,12pt]{article}
\usepackage{graphicx,epsfig}
\usepackage{dcolumn}
\usepackage{fullpage}
\usepackage{appendix}
\usepackage{bm}
\usepackage{amsmath}
\usepackage{amssymb}
\usepackage{amsfonts}
\newcommand{\beq}{\begin{equation}}
\newcommand{\eeq}{\end{equation}}
\newcommand{\bea}{\begin{eqnarray}}
\newcommand{\eea}{\end{eqnarray}}
\newcommand{\beal}{\begin{aligned}}
\newcommand{\eeal}{\end{aligned}}

\renewcommand{\Re}{\widetilde{\operatorname{\mathfrak{Re}}}}
\renewcommand{\Im}{\operatorname{\mathfrak{Im}}}
\newcommand{\pslash} {\ensuremath{\slash \hspace*{-2mm} p}}

\begin{document}
\def\lsim{\:\raisebox{-0.5ex}{$\stackrel{\textstyle<}{\sim}$}\:}
\def\gsim{\:\raisebox{-0.5ex}{$\stackrel{\textstyle>}{\sim}$}\:}

\begin{center}

{\Large \bf On the On-Shell Renormalization of the Chargino and
  Neutralino Masses in the MSSM} \\

\vspace*{1cm}

{}\footnote{Authors' names are listed in alphabetical order}{\sc
  Arindam Chatterjee$^{1}$, Manuel Drees$^{1}$, Suchita
  Kulkarni$^{1}$}\\ and\\ {\sc Qingjun Xu$^{2}$\footnote{On leave from 
    Institute of Nuclear Physics, PAN, Krak\'ow, ul. Radzikowskiego
    152, Poland.}} \\
\vspace*{4mm}
\end{center}
$^{1}$~Bethe Center for Theoretical Physics and Physikalisches
Institut, Universit\"at Bonn, Bonn, 53115, Germany \\ 
$^{2}$~Department of Physics, Hangzhou Normal University, Hangzhou
310036, China

\begin{abstract}
  We discuss the choice of input parameters for the renormalization of
  the chargino and neutralino sector in the minimal supersymmetric
  standard model (MSSM) in the on--shell scheme.  We show that one
  should chose the masses of a bino--like, a wino--like and a
  higgsino--like state as inputs in order to avoid large corrections
  to the masses of the other eigenstates in this sector. We also show
  that schemes where the higgsino--like input state is a neutralino
  are more stable than those where the mass of the higgsino--like
  chargino is used as input. The most stable scheme uses the masses of
  the wino--like chargino as well as the masses of the bino-- and
  higgsino--like neutralinos as inputs.

\end{abstract}

\clearpage

\section{Introduction}
\label{sec:intro}

Among possible solutions of the hierarchy problem
\cite{Witten:1981nf,Sakai:1981gr, Dimopoulos:1981zb, Kaul:1981hi},
supersymmetric extensions of the Standard Model (SM) of particle
physics \cite{Drees2, Baer:2006rs} have the advantage that effects of
the hypothetical new ``superparticles'' required in these
constructions can be computed perturbatively. This distinguishes them
e.g. from scenarios where electroweak symmetry breaking is
accomplished by some new strong dynamics.\footnote{The ``hidden
  sector'' required to break supersymmetry may have non--perturbative
  couplings, but, being hidden, this generally does not invalidate
  perturbative calculations in the visible sector.} At the same time,
future measurements at the LHC \cite{ATLAS_tdr} and a possible
$e^+e^-$ linear collider \cite{LHC/LC} can easily have statistical
uncertainties at or below the percent level, making correspondingly
accurate theoretical calculations highly desirable. This will require
the calculation of electroweak corrections at least at the one--loop
level. Independent motivation for such calculations comes from the
possibility that neutralinos form the dark matter in the
universe, since the determination of the overall dark matter density
of the universe has now been determined with an error of a few percent
\cite{Keisler:2011aw}.

In all these cases the renormalization of the chargino and neutralino
sector plays a central role. In most SUSY breaking scenarios the
lighter charginos and neutralinos are among the lighter
superparticles, i.e. among the first ones to be discovered by future
$e^+e^-$ supercolliders. At the LHC the most important search channels
start from the production of squarks and gluinos, which subsequently
decay into charginos and/or neutralinos. 

The renormalization scheme of choice is on--shell renormalization, for
(at least) two reasons. First, there is a fairly direct relation
between the physical chargino and neutralino masses and the parameters
that need to be renormalized; this relation becomes trivial in the
limit where the spontaneous breaking of the electroweak gauge symmetry
can be neglected. Moreover, masses will probably be among the first
quantities that are measured accurately.

The on--shell renormalization of the neutralino and chargino sectors
in the minimal supersymmetric standard model (MSSM) with conserved
$R-$parity has been discussed in \cite{Fritzsche, Oller, Eberl}. In
full generality, the chargino and neutralino mass matrices depend on
six independent parameters that need to be renormalized. However, two
of these (the masses of the $W$ and $Z$ bosons) already appear in the
SM; they are renormalized through the same conditions as in the SM
(albeit with additional contributions to the relevant two--point
functions with superparticles in the loop). A third parameter, the
ratio of vacuum expectation values (VEVs) $\tan\beta$, is usually
renormalized in the MSSM Higgs sector. This leaves only three
independent parameters whose renormalization has to be fixed through
on--shell conditions for charginos and neutralinos. This means that
only three of the six chargino and neutralino masses can be used as
independent input parameters; their values are exact to all orders in
the on--shell scheme. The other three masses can then be predicted,
but the numerical values of these predictions are subject to loop
corrections.

Previously different calculations used different sets of input masses,
corresponding to different variants of the on--shell scheme.  For
example, in ref.~\cite{Fritzsche}, both chargino masses and the
lightest neutralino mass are used as inputs. In \cite{Drees} the decay
of the next to lightest neutralino to the lightest neutralino was
studied at one loop level using the on--shell renormalization scheme.
The input masses were conveniently assumed to be the masses of the two
lightest neutralinos and the heaviest chargino.

Since the lighter neutralino states are more accessible at colliders,
using their masses as inputs seems favorable from the experimental
point of view. However, reliable perturbative predictions can be made
only if the perturbative expansion is stable, i.e. if loop corrections
are not very large.  From this point of view using the masses of both
charginos and the {\em lightest} neutralino as input can be
problematic: the corrections to the masses of some of the other three
neutralinos can become large if the lightest neutralino is an almost
pure higgsino or wino \cite{Guasch, Baro2}. The reason is that all
three input masses then only depend very weakly on the $U(1)_Y$
gaugino (bino) mass parameter. A large (finite part of the)
counterterm may therefore be needed in order to correct small loop
corrections to the input masses, so that these masses remain
unchanged, as required in on--shell renormalization; the large
counterterm will then lead to large changes of {\em other} neutralino
masses. By a very similar argument, in this situation small
experimental errors on the input masses can lead to very large errors
on the bino mass parameter even when quantum corrections are
neglected. An analogous instability can arise when using the masses of
the two lightest neutralinos and the heavier chargino as inputs. Note
also that both these schemes require the mass of the {\em heavier}
chargino as input, which is one of the heaviest states in this sector;
this considerably weakens the argument in favor of using only the
masses of the lighter neutralino states as inputs.

If the mass of the higgsino--like chargino is used as input an
additional perturbative instability occurs if the $SU(2)$ gaugino
(wino) and higgsino mass parameters are of similar magnitude. It has
been noticed in \cite{Fritzsche, Guasch, Baro2} that using both
chargino masses as input leads to undefined renormalized masses for
exact equality of the absolute values of the wino and higgsino mass
parameters; not surprisingly, corrections remain very large, albeit
finite, if the difference between these two parameters is nonzero but
small. We find that this instability also occurs in schemes where the
mass of the higgsino--like chargino and two neutralino masses are used
as inputs.

In this paper we address these two stability issues by explicitly
calculating chargino and neutralino masses at one--loop level using
the on--shell renormalization scheme. We compare seven different
choices for the three input masses, combining two, one or zero
chargino masses with either the mass(es) of the lightest neutralino
state(s) or with the masses of neutralino states of given character
(bino--, wino-- or higgsino--like). We demonstrate explicitly that the
choice of any three input masses, which include a higgsino--like, a
wino--like and a bino--like mass eigenstate, avoids the first
instability, independent of whether chargino or neutralino masses are
used. Requiring in addition perturbative stability also for similar
magnitudes of the wino and higgsino mass parameters singles out
schemes where the masses of the (most) bino-- and higgsino--like {\em
  neutralinos} are used as inputs. Schemes using three neutralino
masses as inputs show an additional, but much weaker, instability for
approximately equal $SU(2)$ and $U(1)_Y$ gaugino masses; this second
instability is absent if the mass of the wino--like chargino, rather
than that of the wino--like neutralino, is used as input together with
the two neutralino masses listed above.

In this paper we assume all relevant mass parameters to be real.
Experimental constraints on CP violation, mainly from EDM
measurements, require their phases to be small, unless superpartners
of first (and second) generation fermions are very heavy
\cite{Ibrahim, Pospelov}.

This article is organized as follows. In Section~\ref{sec:form} we
discuss briefly seven different variants of the on--shell scheme, in
the context of the MSSM with exact $R-$parity. Section~\ref{sec:prac}
discusses the practical implementation of our schemes. In
Section~\ref{sec:num} we compare the above seven schemes, defined by
different sets of input masses. Explicit expressions for the
counterterms in schemes where two or three neutralino masses are used
as inputs are given in Appendices A and B, respectively, while
Appendix C shows how to distinguish the wino-- and higgsino--like
chargino state using $Z \tilde \chi^+ \tilde \chi^-$ couplings.

\section{Formalism}
\label{sec:form}

In this section we discuss the on--shell renormalization
of the chargino and neutralino sector of the MSSM with exact $R-$parity
\cite {Fritzsche, Baro2}. To this end we first review tree--level
results for chargino and neutralino masses and mixings. We then
describe general features of the on--shell renormalization scheme as
applied to this sector, before introducing three variants of this
scheme, each of which has two or three subvariants.

\subsection{Tree--Level Results}

The tree level mass terms for the charginos, in the gauge eigenbasis,
can be written as \cite{Drees2}
\beq
-\mathcal{L}^{\rm c}_{\rm mass}  =  \psi^{-T} M^c \psi^{+} + h.c.
\eeq
where
\beq 
\psi^+ = (\widetilde{W}^{+},~~\tilde{h}^{+}_{2})^{T},~~ 
\psi^- = (\widetilde{W}^{-},~~\tilde{h}^{-}_{1})^{T}
\eeq
are column vectors whose components are Weyl spinors\footnote{Note
  that $\widetilde W^+$ is the antiparticle of $\widetilde W^-$, but
  $\tilde h^+_2$ is not related to $\tilde h^-_1$; the latter two
  fields reside in the two distinct Higgs superfields required in the
  MSSM.}. The mass matrix $M^{\rm c}$ is given by
\beq \label{mc}
M^{\rm c}= \left( \begin{array}{cc}
M_{2} & \sqrt{2} M_W\sin \beta \\
\sqrt{2} M_W \cos \beta & \mu \\
\end{array} \right).
\eeq
Here $M_2$ is the SUSY breaking $SU(2)$ gaugino (wino) mass, $\mu$ is
the supersymmetric higgsino mass, $M_W$ is the mass of the $W$ boson,
and $\tan\beta$ is the ratio of VEVs of the two neutral Higgs fields
of the MSSM. $M^{\rm c}$ may be diagonalized using unitary matrices
$U $ and $V$ to obtain the diagonal mass matrix,
\beq \label{mcd}
M^{\rm c}_{D} = U^*M^{\rm c}V^{-1} = \left( \begin{array}{cc}
m_{\tilde{\chi}_{1}^{+}} & 0 \\
0 & m_{\tilde{\chi}_{2}^{+}} \\
\end{array} \right). 
\eeq
Without loss of generality, we order the eigenstates such that $0 <
m_{\tilde \chi_{1}^{+}} \leq m_{\tilde \chi_{2}^{+}}$. Note that
scenarios with $m_{\tilde \chi_1^+} \lsim 100$ GeV are excluded by
chargino searches at LEP \cite{PDG}. The left-- and right--handed
components of the corresponding Dirac mass eigenstates, the
charginos $\tilde \chi^+_i$ with $i = 1$ or $2$, are 
\beq \label{ec}
P_L \tilde \chi^{+}_i = V_{ij} \psi^+_j,~~ P_R \tilde \chi^+_i = U^*_{ij}
  \overline{\psi^-_j}\,,
\eeq
where $P_L$ and $P_R$ are chiral projectors, $\overline{\psi^-_j}$ is the
hermitean conjugate of the Weyl fermion $\psi^-_j$, and summation over
$j$ is understood in Eqs.(\ref{ec}).

In the gauge eigenbasis, the tree level neutralino mass terms are
given by \cite{Drees2} 
\beq
-\mathcal{L}^{\rm n}_{\rm mass}  =  \frac{1}{2} \psi^{0T} M^n
\psi^{0} + h.c. , 
\eeq
where
\beq
\psi^{0} = \left( \begin{array}{cccc} \tilde{B^{0}}, & \widetilde W^3, &
    \tilde h_1^0, & \tilde h_2^0 \end{array}\right)^{T}, 
\eeq
is again a column vector whose components are Weyl fermions, and the
neutralino mass matrix $M^{\rm n}$ is given by
\beq \label{mn}
M^{\rm n} =  \left( \begin{array}{cccc}
M_{1} & 0 & -M_{Z}s_W c_\beta & M_{Z}s_W s_\beta\\
0 &  M_{2} & M_{Z}c_W c_\beta & -M_{Z}c_W s_\beta \\
-M_{Z}s_W c_\beta & M_{Z}c_W c_\beta & 0 & -\mu \\ 
M_{Z}s_W s_\beta & -M_{Z}c_W s_\beta & -\mu & 0 \end{array} \right)\,.
\eeq
Here $s_W, s_\beta, c_W$ and $c_\beta$ stand for $\sin \theta_W, \sin
\beta, \cos \theta_W$ and $\cos \beta$, respectively, where $\theta_W$
is the weak mixing angle. $M_Z$ is the mass of the $Z$ boson, and
$M_1$ is the SUSY breaking $U(1)_Y$ gaugino (bino) mass. Finally,
$M_2$ and $\mu$ already appeared in the chargino mass matrix
(\ref{mc}).

$M^{\rm n}$ may be diagonalized by a unitary matrix $N$ to obtain the
diagonalized mass matrix $M^{\rm n}_{D}$,
\beq \label{mnd}
M^{\rm n}_{D}=N^* M^{\rm n} N^{-1}=\left( \begin{array}{cccc}
m_{\tilde{\chi}^{0}_{1}} & 0 & 0 & 0\\
0 &  m_{\tilde{\chi}^{0}_{2}} & 0 & 0 \\
0 & 0 & m_{\tilde{\chi}^{0}_{3}} & 0 \\ 
0 & 0 & 0 & m_{\tilde{\chi}^{0}_{4}} \end{array} \right).
\eeq
Again, without loss of generality, we order the eigenvalues such that
$$ 0 \leq m_{\tilde{\chi}^{0}_{1}} \leq m_{\tilde{\chi}^{0}_{2}} \leq
m_{\tilde{\chi}^{0}_{3}} \leq m_{\tilde{\chi}^{0}_{4}}. $$ Note that
an arbitrarily light neutralino is still phenomenologically possible,
if no relations between $M_1, \, M_2$ and/or $\mu$ are imposed
\cite{Dreiner:2009ic}.

The left--handed components of the corresponding mass eigenstates,
described by four--component Majorana neutralinos $\tilde \chi_i^0$
with $i \in \{1,2,3,4\}$, may be obtained as,
\beq
P_L \tilde \chi^{0}_i = N_{ij}  \psi_j^{0},
\eeq
where summation over $j$ is again implied; the right--handed
components of the neutralinos are determined by the Majorana condition
$\tilde \chi_i^C = \tilde \chi_i$, where the superscript $C$ stands
for charge conjugation.

The parameters $M_2$ and $\beta$ can always be made real and positive
through phase rotations of fields. If the parameters $M_1$ and $\mu$
are also real, as we assume, the chargino mixing matrices $U$ and $V$
can be chosen to be purely real. However, the neutralino mixing matrix
$N$ can only chosen to be real if negative neutralino ``masses''
(better: eigenvalues of $M^{\rm n}$) are tolerated. Physical masses
must not be negative, of course. In that case the entries of $N$ can
be either real or purely imaginary. Genuinely complex entries of $N$,
as well as $U$ and $V$, are only required if $M_1$ and/or $\mu$ have
non--trivial CP--odd phases.

\subsection{Generalities of On--Shell Renormalization}

At the quantum level, diagrams with internal and/or external chargino
and/or neutralino lines will receive two--point function corrections
on all these lines. These corrections are in general infinite. In
order to absorb such infinities through renormalization, the mass
matrices and the mass eigenstates need to be redefined. Here we
discuss this renormalization at the one--loop level.

The one--loop mass eigenstates can be related to the tree level mass
eigenstates through wave function renormalization:
\beq \label{eZ}
\tilde{\chi_{i}}^{\rm bare} = (\delta_{ij}+\frac{1}{2} \delta Z_{ij} P_{L} +
\frac{1}{2} \delta Z_{ij}^{*} P_{R})~ \tilde{\chi_{j}}^{\rm renormalized}.
\eeq 
This relation holds for both the chargino sector, with $\tilde \chi_i
\equiv \tilde \chi_i^+, \, i \in \{1,2\}$, and the neutralino sector,
with $\tilde \chi_i \equiv \tilde \chi_i^0, \, i \in \{1,2,3,4\}$.
Note that the wave function renormalization matrices $\delta Z^{\rm
  c}$ and $\delta Z^{\rm n}$ in the chargino and neutralino sectors
are not diagonal. In the on--shell renormalization scheme
loop--induced changes of the identities of the mass eigenstates can be
described entirely through these wave function renormalizations. The
entries of the mixing matrices $N, U$ and $V$ can therefore be chosen
to be identical to their tree level values \cite{Baro2}. Explicit
expressions for the wave function renormalization constants are not
needed for the calculation of the chargino and neutralino masses, on
which we focus in this paper.

These masses receive explicit corrections from one--loop diagrams,
which can be written as
\beq \label{dmf}
\delta m_{f} = \frac{1}{2} m_{f} \left[ \Re\Sigma^{VL}_{ff}(m_{f}^{2}) +\Re 
  \Sigma^{VR}_{ff}(m_{f}^{2}) \right] + \frac{1}{2} \left[ \Re
  \Sigma^{SL}_{ff}(m_{f}^{2}) + \Re \Sigma^{SR}_{ff}(m_{f}^{2}) \right], 
\eeq
where $f$ denotes a fermion species with mass $m_f$. $\Re$ denotes the
real parts of the loop integrals involved, leaving imaginary parts of
couplings unchanged (in the CP--conserving case these can only appear
through the neutralino mixing matrix $N$, as remarked at the end of
the previous Subsection). Moreover, the $\Sigma$ refer to various
terms in the general two--point function of fermion $f$ in momentum
space:
\beq \label{Sigma}
\Sigma_{ff}(p) = \pslash \left[ P_L \Sigma^{VL}(p) + P_R
  \Sigma^{VR}(p) \right] + P_L \Sigma^{SL}(p) + P_R \Sigma^{SR}(p)\,,
\eeq
where $P_L$ and $P_R$ are again the chiral projectors. Note that only
diagonal two--point functions contribute to the corrections to
physical masses at one--loop level in the on--shell
scheme.\footnote{Off--diagonal two--point function corrections do
  appear in generic one--loop diagrams. Their infinities are absorbed
  in off--diagonal wave function renormalization constants, which are
  fixed by the definition that particles do not mix on--shell. See
  e.g. ref.\cite{Fritzsche} for further details.}

The corrections of Eq.(\ref{dmf}) are in general divergent. These
divergencies are absorbed into counterterms to the mass matrices
$M^{\rm c}$ and $M^{\rm n}$ of Eqs.(\ref{mc}) and (\ref{mn}),
respectively, i.e.
\beq \label{delM}
M^{\rm bare} = M^{\rm renormalized} + \delta M.
\eeq
As usual, we define the physical (on--shell) masses as poles of the
real parts of the (one--loop corrected) propagators. The physical
chargino masses are then given by,
\beq \label{m_pchar}
m^{\rm os}_{\tilde\chi_{i}^{+}} = m_{\tilde\chi_{i}^{+}} + (U^* \delta
M^{\rm c}V^{-1})_{ii} - \delta m_{\tilde\chi_{i}^{+}}\,;
\eeq
the corresponding expression for the neutralinos is 
\beq \label{m_pneut}
m^{\rm os}_{\tilde\chi_{i}^{0}} = m_{\tilde\chi_{i}^{0}} + (N^*\delta
M^{\rm n}N^{-1})_{ii} - \delta m_{\tilde\chi_{i}^{0}} \,.
\eeq
The masses $m_{\tilde\chi_i^{+,0}}$ appearing on the right--hand sides
of Eqs.(\ref{m_pchar}) and (\ref{m_pneut}) are the (finite)
tree--level masses. $U, V$ and $N$ are the mixing matrices in the
chargino and neutralino sectors, respectively, which, as already
noted, do not get modified by loop corrections. $\delta m_{\tilde
  \chi_i^{+,0}}$ are the explicit loop corrections of Eq.(\ref{dmf})
as applied to the charginos and neutralinos. Finally, $\delta M^{\rm
  c, n}$ are the counterterm matrices of Eq.(\ref{delM}) for the
chargino and neutralino sector, which we yet have to determine.

Renormalizability requires that the counterterm matrices have exactly
the same form as the tree--level mass matrices, i.e. all
non--vanishing entries are replaced by the corresponding counterterms
while the vanishing entries receive no correction. To one--loop order
counterterms to products like $M_W \sin\beta$ can be written as
$(\delta M_W) \sin\beta + M_W \delta \sin\beta$, and so on. 

Altogether there are thus seven different counterterms: $\delta M_W,\,
\delta M_Z,\, \delta \theta_W, \, \delta \tan\beta, \, \delta M_1$, 
$\delta M_2$ and $\delta \mu$. The first three of these already appear
in the SM. We renormalize them according to the on--shell prescription
of electroweak renormalization, where $M_W$ and $M_Z$ are physical
(pole) masses, and $\cos\theta_W = M_W/M_Z$. This gives
\cite{Marciano:1980pb}: 
\bea \label{delta_SM}
\delta M_{W}^{2} & = & \Re\Sigma_{WW}(M_{W}^{2})\,; \nonumber \\
\delta M_{Z}^{2} & = & \Re\Sigma_{ZZ}(M_{Z}^{2})\,; \nonumber \\
\delta \cos \theta_W & = & \frac{M_W}{M_Z} \left(\frac{\delta M_W}{M_W}-
  \frac{\delta M_Z}{M_Z} \right)\,.
\eea
Here $\Sigma_{WW}$ and $\Sigma_{ZZ}$ are the transverse components of
the diagonal $W$ and $Z$ two--point functions in momentum space,
respectively; again only the real parts of the loop functions should
be included as indicated by the $\Re$ symbols. Note that, while the
definitions of these three counterterms are formally as in the SM, in
the MSSM there are many new contributions to $\Sigma_{WW}$ and
$\Sigma_{ZZ}$ involving loops of superparticles and additional Higgs
bosons.

\setcounter{footnote}{0}

The counterterms (\ref{delta_SM}) imply that the sum of the squares of
the VEVs of the two neutral Higgs components is already
fixed. However, the ratio of these VEVs, $\tan\beta$, does not affect
the $W$ and $Z$ masses, hence its counterterm has not yet been
determined. Following refs.~\cite{Chankowski, Dabelstein} we fix it by
the requirement that the two--point function connecting the CP--odd Higgs
boson $A$ to the (longitudinal part of the) $Z-$boson vanishes when
$A$ is on--shell. This gives\footnote{Note that the couplings of $A$
  contain an extra factor of $i$ relative to gauge couplings.
  Therefore the imaginary part of the two--point function appears in
  Eq.(\ref{delta_tanbeta}); this contains the real (dispersive),
  infinite, part of the loop function.}
\beq \label{delta_tanbeta}
\delta \tan \beta = \frac{1}{2 M_Z \cos^{2} \beta }
\Im(\Sigma_{AZ}(m_{A}^{2}))\,.
\eeq

Hence only the counterterms to $M_1, \, M_2$ and $\mu$ remain to be
fixed in the chargino and neutralino sector. This means that we can
only require that three of the six physical masses in this sector are
not changed by loop corrections, i.e. only three of the six
``tree--level'' masses are physical (all--order) masses. The other
three masses will receive finite, but non--zero
corrections.\footnote{The finiteness of these corrections is a
  condition for the renormalizability of the theory. In practice it
  affords non--trivial checks of our calculation.} In order to
complete the definition of our renormalization scheme, we have to
decide which three masses to use as input masses. There are many ways
to do so; this is the topic of the following Subsection.

\subsection{Choosing Input Masses}
\label{sec:schemes}

For a fixed point in parameter space, there are $\left( \begin{array}{c}
    6 \\ 3 \end{array} \right) = 20$ different ways to select three out
of six masses. However, as we will see shortly, {\em any} scheme that
always chooses masses with fixed subscripts $i,j,k$ as inputs is bound
to lead to perturbative instability in large regions of parameter
space. Instead, one should choose the masses of chargino and/or
neutralino states with specific properties as inputs; the indices of
the input states will then vary over parameter space.

We illustrate this by considering three different schemes, each of
which has two or three variants. The most widely used inputs are the
masses of both charginos and of the lightest neutralino
\cite{Fritzsche, Baro2}. We call this {\bf scheme 1a}. Using both
chargino masses as inputs guarantees that the finite parts of the
counterterms $\delta M_2$ and $\delta \mu$ are usually small, since at
least one chargino mass will have strong sensitivity to either of
these counterterms. However, as already mentioned in the Introduction,
this choice leads to instabilities if $|N_{11}| \ll 1$ \cite{Guasch,
  Baro2}. In this case the tree--level value of $m_{\tilde \chi_1^0}$
depends only very weakly on $M_1$. Correspondingly the counterterm
$\delta M_1$ may require a very large finite part to cancel the finite
part of the explicit corrections to $m_{\tilde \chi_1^0}$ in
Eq.(\ref{m_pneut}). This in turn will lead to a very large correction
to the mass of the (most) bino--like
neutralino.\footnote{Symbolically, if $\partial m_{\tilde
    \chi_1^0}^{\rm tree} / \partial M_1 = \epsilon$, and explicit
  finite loop corrections to $m_{\tilde \chi_1^0}$ are of typical
  (relative) one--loop order, i.e. ${\cal O}(\alpha/ \pi)$, then the
  finite part of the counterterm to $M_1$ will be of relative order
  ${\cal O}(\alpha/(\epsilon \pi))$, where $\alpha$ is an electroweak
  fine structure constant. The finite correction to the mass of the
  most bino--like neutralino will then also be enhanced by
  $1/\epsilon$, and will thus become very large if $\epsilon \lsim
  \alpha$.} In terms of parameter space, this instability will occur
whenever $|M_1| - M_2 \gg M_Z$ {\em or} $|M_1| - |\mu| \gg M_Z$, in
which case $\tilde \chi_1^0$ is wino-- or higgsino--like.

A suitable alternative is to instead use the mass of the bino--like
neutralino as an input \cite{Guasch, Baro2}. We call this {\bf scheme
  1b}. In the limit of small mixing, this mass is approximately given
by $|M_1|$, and it will retain strong sensitivity to $M_1$ even in the
presence of significant mixing.

As already mentioned, in the on--shell scheme the input masses are
interpreted as exact physical masses, i.e. their total one--loop
corrections should vanish. Eqs.(\ref{m_pchar}) and (\ref{m_pneut})
then lead to the following equations:
\bea \label{scheme1ab}
(U^*\delta M^cV^{-1})_{11} & = & \delta m_{\tilde\chi_{1}^{+}}, 
\nonumber \\
(U^*\delta M^cV^{-1})_{22} & = &\delta m_{\tilde\chi_{2}^{+}}, 
\nonumber \\
(N^*\delta M^nN^{-1})_{ii} & = & \delta m_{\tilde\chi_{i}^{0}}
~\text{(no summation)}, 
\eea
where $i=1$ in {\bf scheme 1a}, while $i$ denotes the (most)
bino--like neutralino in {\bf scheme 1b}. Note that both chargino
equations in (\ref{scheme1ab}) depend on both $\delta \mu$ and $\delta
M_2$, as well as on $\delta M_W$ and $\delta \tan\beta$ which have
already been determined in Eqs.(\ref{delta_SM}) and
(\ref{delta_tanbeta}), respectively. Similarly, the third (neutralino)
equation in (\ref{scheme1ab}) depends on all the
counterterms. However, since the dependence on counterterms is linear,
eqs.(\ref{scheme1ab}) can readily be solved analytically.

Alternatively, one can use the masses of one chargino and two
neutralinos inputs. The counterterms $\delta M_1, \delta M_2$ and
$\delta \mu$ are then determined by the following equations:
\bea \label{scheme2ab}
(N^*\delta M^nN^{-1})_{ii} & = & \delta m_{\tilde\chi_{i}^{0}},
\nonumber  \\ 
(N^*\delta M^nN^{-1})_{jj} & = & \delta m_{\tilde\chi_{j}^{0}},
\nonumber  \\ 
(U^*\delta M^cV^{-1})_{kk} & = &\delta m_{\tilde\chi_{k}^{+}} ~
\text{(no summation)}.
\eea
In {\bf scheme 2a} we take $i = 1, j = 2$ and $k = 1$, i.e. the
lightest chargino mass and the two lightest neutralino masses are used
as inputs. These are the lightest particles in this sector, hence
their masses are likely to be the first to be determined
experimentally. However, this scheme is likely to lead to perturbative
instabilities whenever the differences between $|M_1|, \, M_2$ and
$|\mu|$ are large. For example, if $M_2 - |\mu| \gg M_Z$, $\tilde
\chi_1^+$ and $\tilde \chi_2^0$ will both be higgsino--like, and none
of the input masses will be sensitive to $M_2$, leading to a
potentially large finite part of $\delta M_2$. If $|\mu| - M_2 \gg
M_Z$ and $|\mu| - |M_1| \gg M_Z$, none of the input states is higgsino
like; hence none of the input masses depends sensitively on $\mu$,
leading to a potentially large finite part of $\delta \mu$. In {\bf
  scheme 2b} $i$ and $j$ therefore denote the bino-- and
wino--like neutralino, respectively, while $k$ denotes the
higgsino--like chargino. This ensures that there is at least one input
mass that is sensitive to each counterterm. This is true also in {\bf
  scheme 2c}, where $i$ and $j$ denote the bino-- and higgsino--like
neutralino, respectively, while $k$ denotes the wino--like chargino.

With three neutralino masses as inputs, the counterterms $\delta M_1,
\, \delta M_2$ and $\delta \mu$ are all determined from Eq.(\ref{m_pneut}):
\bea \label{scheme3ab}
(N^*\delta M^nN^{-1})_{ii} & = & \delta m_{\tilde\chi_{i}^{0}},
\nonumber \\ 
(N^*\delta M^nN^{-1})_{jj} & = & \delta m_{\tilde\chi_{j}^{0}} ,
\nonumber \\ 
(N^*\delta M^nN^{-1})_{kk} & = & \delta m_{\tilde\chi_{k}^{0}} ~
\text{(no summation)}. 
\eea
In {\bf scheme 3a} we simply take the three lightest neutralinos as
input states, i.e. $i = 1,\, j = 2$ and $k = 3$ in
Eqs.(\ref{scheme3ab}). This runs the risk that two of the input states
are higgsino--like, in which case the third input mass cannot
determine both $\delta M_1$ and $\delta M_2$ reliably. For example, if
$M_2 - |M_1| \gg M_Z$ and $M_2 - |\mu| \gg M_Z$, none of the input
masses is sensitive to $\delta M_2$, since the wino--like neutralino
is the heaviest one. In {\bf scheme 3b} $i,\, j$ and $k$ therefore
denote the bino--like, wino--like and higgsino--like neutralino,
respectively.

For each set of input masses, the corresponding set of linear
equations can be solved to obtain $\delta M_1$, $\delta M_2$ and
$\delta \mu$. The solutions, in the context of scheme 1a, have been
obtained using the mixing matrices in \cite{Fritzsche} and without
using the mixing matrices in \cite{Guasch, Baro2}. The same solution
can be used for scheme 1b, with the substitution $N_{1\alpha}
\rightarrow N_{i\alpha}$, where $i$ denotes the bino--like
neutralino. Explicit solutions for schemes 2 and 3 can be found in
Appendices A and B, respectively.

The seven schemes are summarized in Table~\ref{schemes}. The second
column lists the states whose masses are used as inputs. Here $\tilde
\chi_b, \, \tilde \chi_w$ and $\tilde \chi_h$ stand for a bino--,
wino-- and higgsino--like state, respectively. The third column lists
regions of parameter space, defined through strong inequalities, where
at least one counterterm is poorly determined, leading to potentially
large corrections to some mass(es); the poorly determined counterterm
is given in parentheses. These regions only exist for schemes 1a, 2a
and 3a; in fact, one can easily convince oneself that such regions of
potential perturbative instability exist in all schemes that fix the
indices of the input states a priori, independent of the characters of
these states.

\begin{table}
\begin{center}
\begin{tabular}{|l|l|l|}
\hline
Scheme & Input states & Regions of instability (counterterm) \\
\hline \hline
1a & $\tilde \chi_1^+,\, \tilde \chi_2^+,\, \tilde \chi_1^0$ &
$|M_1| - |\mu| \gg M_Z \ (\delta M_1)$ \\
& & $|M_1| - M_2 \gg M_Z \ (\delta M_1)$ \\
\hline
1b & $\tilde \chi_1^+, \, \tilde \chi_2^+, \, \tilde \chi_b^0$ & \\
\hline \hline
& & $M_2 - |\mu| \gg M_Z \ (\delta M_2)$ \\
2a & $\tilde \chi_1^+, \, \tilde \chi_1^0, \, \tilde \chi_2^0$ & 
$|\mu| - M_2 \gg M_Z$ and $|\mu| - |M_1| \gg M_Z \ (\delta \mu)$ \\
& & $|M_1| - |\mu| \gg M_Z \ (\delta M_1)$ \\
\hline
2b & $\tilde \chi_h^+, \, \tilde \chi^0_b, \, \tilde \chi^0_w$ & \\
\hline
2c & $\tilde \chi_w^+, \, \tilde \chi^0_b, \, \tilde \chi^0_h$ & \\
\hline \hline
& & $M_2 - |\mu| \gg M_Z$ and $|M_1| - |\mu| \gg M_Z \ (\delta M_1$ or
$\delta M_2$) \\
3a & $\tilde \chi_1^0,\, \tilde \chi_2^0,\, \tilde \chi_3^0$ &
$M_2 - |\mu| \gg M_Z$ and $|M_1| \lsim |\mu| \ (\delta M_2)$ \\
& & $|M_1| - |\mu| \gg M_Z$ and $M_2 \lsim |\mu| \ (\delta M_1)$ \\
\hline
3b & $\tilde \chi^0_b,\, \tilde \chi_w^0,\, \tilde \chi_h^0$ & \\
\hline
\end{tabular}
\end{center}
\caption{Summary of the schemes discussed in this paper. The second
  column lists the states whose masses are used as inputs; here
  $\tilde \chi_b, \, \tilde \chi_w$ and $\tilde \chi_h$ stands for a
  bino--, wino-- and higgsino--like state, respectively. The last
  column lists the regions of parameter space where schemes 1a, 2a and
  3a may become perturbatively unstable, together with the counterterm
  that is poorly determined in that region. In the first instability
  region of scheme 3a, only one linear combination of $\delta M_1$ and
  $\delta M_2$ is well determined, depending on the relative ordering
  of $|M_1|$ and $M_2$. Note that the different regions of instability
may overlap. }
\label{schemes}
\end{table}

This concludes the description of the renormalization schemes we are
using. Before we turn to numerical results, we briefly discuss some
issues that might arise in the practical implementation of our formalism.

\section{Practical Considerations}
\label{sec:prac}

In this Section we discuss some issues regarding the practical
implementation of our calculation. In particular, we define the
bino--, wino-- and higgsino--like states in terms of observable
quantities. We then discuss how to extract the on--shell parameters
$M_1, \, M_2$ and $\mu$ from the three input masses in the three
schemes. Finally, we describe how to combine our formalism with
spectrum calculators.

\subsection{Definition of the States}
\label{subsec:states}

We saw at the end of the previous Section that schemes where the three
states whose masses are taken as inputs include a bino--like, a
wino--like and a higgsino--like state are better behaved than schemes
where the indices of the input states are fixed a priori. This raises
the question how these input states are defined.

We begin with the neutralino sector. The bino--like state is defined
as the neutralino $\tilde \chi_i^0$ with the largest bino component,
i.e. $|N_{i1}| \geq |N_{j1}|\ \forall j \neq i$. This can (at least in
principle) be determined experimentally by finding the neutralino with
the strongest coupling to right--handed electrons, i.e. with largest
absolute value of the $\tilde \chi_1^0 e_R \tilde e_R$
coupling.\footnote{Occasionally there will be two neutralinos $\tilde
  \chi_i^0, \, \tilde \chi_j^0$ with $|N_{i1}| \simeq |N_{j1}| \simeq
  1/\sqrt{2}$. In this case it doesn't matter whether $m_{\tilde
    \chi_i^0}$ or $m_{\tilde \chi_j^0}$ is chosen as input, since both
  masses are then quite sensitive to $M_1$.}

The (most) wino--like neutralino is the state $\tilde \chi_i^0$ with
largest wino component, i.e. $|N_{i2}| \geq |N_{j1}|\ \forall j \neq
i$. Having already determined the bino--like neutralino, the
wino--like state is the remaining neutralino with the strongest
coupling to left--handed sleptons, i.e. the (non--bino) state with
largest absolute values of the $\tilde \chi_i^0 e_L \tilde e_L$ and
$\tilde \chi_i^0 \nu_L \tilde \nu_L$ couplings.\footnote{Occasionally
  the state $\tilde \chi_i^0$ with the largest $|N_{i1}|$ {\em also}
  has the largest $|N_{i2}|$, i.e. the most bino--like state is also
  the most wino--like of all four states; this can happen in the
  presence of strong bino--wino mixing, i.e. $M_1 \simeq M_2$, if in
  addition $||\mu| - M_2| \lsim M_Z$. This ambiguity is resolved by
  first determining the most bino--like of all four neutralinos, and
  then determining the most wino--like of the remaining three
  neutralinos. A similar situation may be encountered in case of 
  strong bino--higgsino mixing, especially when $M_2 \simeq |\mu|$.}

Finally, we define the (most) higgsino--like state as the one with
largest $(|N_{i3}|^2 +|N_{i4}|^2)$. Unfortunately this sum does not
directly correspond to a measurable coupling. However, using unitarity
of the mixing matrix we can write $|N_{i3}|^2 + |N_{i4}|^2 = 1 -
|N_{i1}|^2 - |N_{i2}|^2$, i.e. the most higgsino--like state is the
one with weakest couplings to first generation leptons. Moreover, once
the bino-- and wino--like states have been identified, in most cases
it does not matter very much which of the two remaining states is
defined to be ``the'' higgsino--like state.\footnote{We will see in
  the next Section that the definition of the higgsino--like
  neutralino does matter if $M_2 \simeq |\mu|$ or $-M_1 \simeq
  |\mu|$. The definition we employ here maintains perturbative
  stability also in these scenarios with strong mixing.}

In the chargino sector, one can define the wino--like state as the
state which has larger coupling to a (s)neutrino or to a left--handed
(s)electron; the other state is then higgsino--like. Alternatively, one
can take the strength of the $\tilde \chi_i^+ \tilde \chi_i^- Z^0$
coupling as defining property: since the wino is an $SU(2)$ triplet
while the higgsino is a doublet, the former couples more strongly to
the $Z$. We show in Appendix C that this also allows a unique
definition of the wino-- and higgsino--like states.

\subsection{Extraction of the On--Shell Parameters}
\label{subsec:parameters}
\setcounter{footnote}{0}

Experimentally one will (hopefully eventually) measure several
neutralino and chargino masses. In order to apply our formalism, one
needs to calculate the input values of $M_1, \, M_2$ and $\mu$ from
the three input masses. This is necessary in order to predict the
other three chargino and neutralino masses, which were not used as
inputs. This inversion from masses to parameters is also required when
comparing results of different schemes (see below).

In schemes 1a and 1b this inversion can be done analytically
\cite{Fritzsche}. Here both chargino masses are inputs. The chargino
mass matrix can easily be diagonalized analytically; this can be
inverted to derive analytical expressions for $M_2$ and $\mu$ in terms
of the chargino masses (and $\tan\beta$, which we assume to be known
independently). In general there are four solutions for $M_2$ and
$\mu$. To begin with, the eigenvalue equations (of $M^{\rm c}M^{\rm c
  \dagger}$ or $M^{\rm c \dagger}M^{\rm c}$) are symmetric under $M_2
\leftrightarrow \mu$. This degeneracy can be lifted once we know
whether the lighter or the heavier chargino is (more) wino like; this
determines whether $M_2$ is smaller or larger than $|\mu|$. Notice
that even in this scheme knowledge of the qualitative properties of
the chargino states is required to complete the inversion. The second
degeneracy occurs since the sign of $\mu$ has not been fixed; recall
that we define $M_2$ to be real and positive. This degeneracy can be
lifted by measuring some chargino coupling or cross section
\cite{Desch} in addition to the masses, since the two solutions will
have different mixing matrices $U$ and $V$. Alternatively one can use
information from the neutralino sector to lift this degeneracy, since
the two solutions will lead to somewhat different neutralino
spectra. To this end, one can simply try both solutions, and see which
one more accurately reproduces the measured neutralino spectrum.

Having determined $M_2$ and $\mu$ from the chargino sector, the
remaining parameter $M_1$ can be determined analytically from the
neutralino sector \cite{Fritzsche}. Again, there is a two--fold
degeneracy, having to do with the sign of $M_1$. This degeneracy would
not occur if we were able to determine the eigenvalue of the
neutralino mass matrix (\ref{mn}) rather than only its absolute value,
which is the physical mass. In fact, the {\em relative} sign between
two eigenvalues, with indices $i$ and $j$, of this matrix {\em is}
physical \cite{Drees2}: if the relative sign is positive, the $Z
\tilde \chi_i^0 \tilde \chi_j^0$ coupling will be purely axial vector,
whereas for negative relative sign it will be purely vector. This can
be determined through $e^+e^- \rightarrow \tilde \chi_i^0 \tilde
\chi_j^0$, where a vector (axial vector) coupling leads to an $S- \
(P-)$wave behavior of the cross section, i.e. to a suppression of the
cross section by one (three) factors of the neutralino
center--of--mass three--momentum.

In the remaining schemes an analytical inversion from input masses to
input parameters is not practical. A numerical solution of the
inversion equations is straightforward, though. Again these equations
will have several solutions, which differ by the signs of $M_1$ and/or
$\mu$, and/or by the ordering of the absolute values of $M_1$, $M_2$
and $\mu$. The latter degeneracy can again be lifted if information
about the characters of the states is available. For example, $|M_1|$
will generally be smaller than $M_2$ if the bino--like neutralino is
lighter than the wino--like one, and so on. The sign issue can be
resolved as in scheme 1.

Note that in this manner one determines $M_1, \, M_2$ and $\mu$ in an
on--shell scheme, as required for our calculation. In order to relate
these to $\overline{\rm DR}$ parameters, one needs information about
the entire superparticle and Higgs spectrum \cite{Pierce:1996zz}.

\subsection{Combination with Spectrum Calculators}
\label{subsec:codes}

Predictions for the masses of superparticles and Higgs bosons for
given values of the relevant parameters of the MSSM Lagrangian are
nowadays usually obtained with the help of publicly available programs
like ISASUSY \cite{ISASUSY}, SuSpect \cite{SUSPECT}, SOFTSUSY
\cite{SOFTSUSY} and SPHENO \cite{SPHENO}. As per the SUSY Les Houches
Accords (SLHA) \cite{Skands:2003cj, Allanach:2008qq} and the
Supersymmetric Parameter Analysis (SPA) convention \cite{SPA}, values
for parameters in the Lagrangian are given in the $\overline{\rm DR}$
scheme, not in the on--shell scheme. This is true also for the
``weak--scale'' output parameters $M_1,\, M_2$ and $\mu$ of the
spectrum calculators (which are typically defined at a SUSY breaking
scale rather than the weak scale). These parameters should {\em not}
be used as inputs in the mass matrices (\ref{mc}) and (\ref{mn}),
since our input parameters are on--shell parameters; that is, in the
limit of no mixing, where all differences between $|M_1|, \, M_2$ and
$|\mu|$ are much bigger than $M_Z$, these parameters become physical
(on--shell) masses in our scheme, which is not the case for the
running ($\overline{\rm DR}$) parameters.

Fortunately modern spectrum calculators already perform the
transformation from $\overline{\rm DR}$ to on--shell parameters
internally, in the limit of no mixing in the neutralino and chargino
sector. Hence one can identify our on--shell input parameters $M_1, \,
M_2$ and $-\mu$ with whatever the spectrum calculators input as the
$(1,1), \, (2,2)$ and $(3,4)$ entries of the neutralino mass matrix,
respectively. Alternatively one can perform the conversion oneself,
using the results of ref.~\cite{Pierce:1996zz}.

However, spectrum calculators generally do not use on--shell $W$ and
$Z$ masses in the off--diagonal elements of the chargino and
neutralino mass matrices. In fact, physically there is no reason why
on--shell masses should be preferable here. We have used them since
the on--shell renormalization of the electroweak sector is very well
established, and since we do not expect large radiative corrections
from these sources anyway. Similarly, spectrum calculators do not
enforce $\cos \theta_W = M_W / M_Z$, which in our scheme is necessary
since otherwise some divergencies remain. The off--diagonal entries of
the chargino and neutralino mass matrices [other than the $(3,4)$ and
$(4,3)$ entries of the latter] should therefore {\em not} be copied
from the spectrum calculators, but have to be entered ``by hand''
following Eqs.(\ref{mc}) and (\ref{mn}). For this reason, the mixing
matrices $U, \, V$ and $N$ should not be taken from the spectrum
calculators, either; rather, they should be calculated by explicitly
diagonalizing the chargino and mass matrices, constructed as per the
above description, e.g. by using some standard matrix diagonalization
routines.

\section{Numerical Results}
\label{sec:num}

In this section we numerically compute the physical masses of the
charginos and neutralinos, given by Eqs.(\ref{m_pchar}) and
(\ref{m_pneut}). We have used FeynArts \cite{Hahn1, Hahn1a}, FormCalc
\cite{Hahn2} and LoopTools \cite{Hahn3} for the computations of the
relevant two--point functions. Feynman gauge has been used
throughout. For regularization we have used the constrained
differential renormalization method \cite{CDR}. At one loop level this
method has been proved to be equivalent \cite{Hahn3} to regularization
by the dimensional reduction method \cite{DR}. All numerical examples
presented below employ the following parameters:
$$
A = 400~ \text{GeV},\, M_{\rm SUSY} = 450 ~\text{GeV},\, \tan \beta =5
,\, M_{A} = 250 ~\text{GeV}. $$ For simplicity we took the $A$
parameters and the supersymmetry breaking soft sfermion masses $M_{\rm
  SUSY}$ to be universal as well as flavor conserving. 

In the first Subsection we survey the parameter space by scanning
$\mu$ or $M_1$. We will see that some counterterms can become
singular, leading to ill--defined one--loop masses at the pole and, in
some cases, to very large corrections near the pole. This issue is
discussed in depth in the second Subsection. The final Subsection
contains numerical comparisons between schemes for a couple of
benchmark points.

\subsection{Surveys of Parameter Space}

In the following figures we show the differences between one--loop and
tree--level masses in percent, defined by
\bea
\Delta^{0}_{i} & = &
\dfrac{m^{\rm 1-loop, \,os}_{\tilde\chi_i^0} - m^{\rm
    tree}_{\tilde\chi_i^0} }
{m^{\rm tree}_{\tilde\chi_i^0} }\times 100, \ \ \ i \in
\{1,2,3,4\}
\nonumber \\
\Delta^{+}_{i} & = &
\dfrac{m^{\rm 1-loop, \, os}_{\tilde\chi_i^+} - m^{\rm
    tree}_{\tilde\chi_i^+}} {m^{\rm
    tree}_{\tilde\chi_i^+} } \times 100, \ \ \ i \in \{1,2\}\,.
\eea
Of course, these differences vanish for the three input states.
We have fixed $M_1 = 89.29 ~ \text{GeV}, ~ M_2 = 369.78
~\text{GeV}$ in all the plots against $\mu$, and $M_2 = 200
~\text{GeV}, ~ \mu = 300 ~\text{GeV}$ in all the plots against $M_1$.

\begin{figure}[t]
\epsfig{file=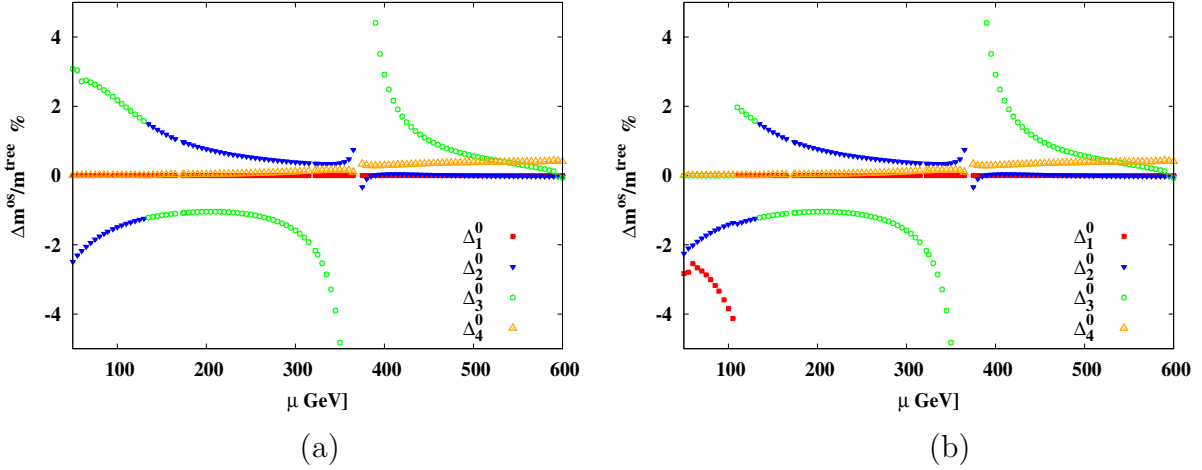, width=\linewidth, height=6 cm}
$~~~~~~~~~~~~~~~~~~~~~~~~~~~~~~ \text{(a)}  \hspace{7.75cm} \text{(b)}$
\caption{Radiative corrections (in \%) to all the neutralino
  masses against $\mu$. The masses of two charginos and one neutralino
have been used as inputs. Frame (a) is for scheme 1a, where the mass
of the lightest neutralino is used as input, whereas frame (b) is for
scheme 1b, where instead the mass of the most bino--like neutralino
is used as input.} 
\label{fs1mu}
\end{figure}

\begin{figure}[!ht]
\epsfig{file=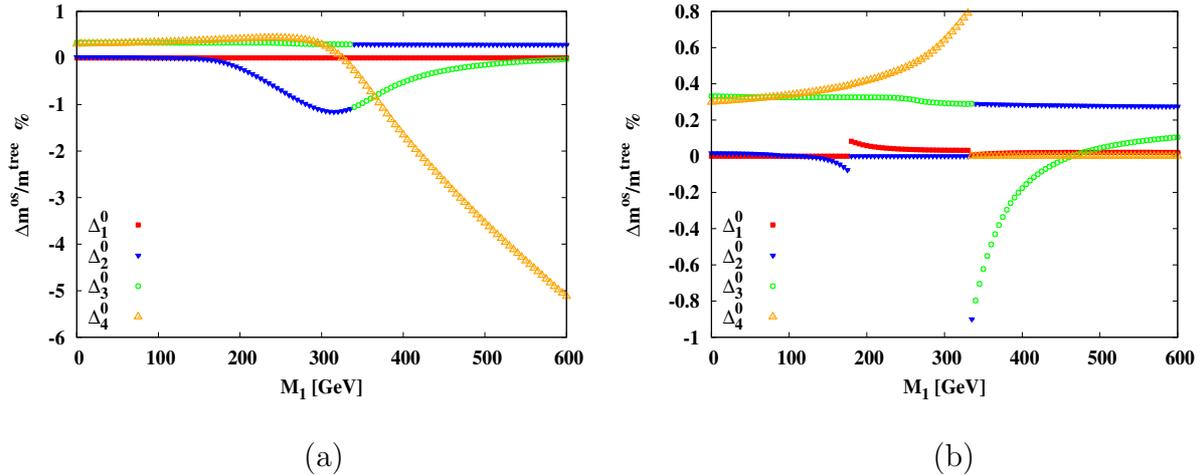,width=\linewidth, height=6 cm} 
$~~~~~~~~~~~~~~~~~~~~~~~~~~~~~~ \text{(a)}  \hspace{7.75cm} \text{(b)}$
\caption{Radiative corrections (in \%) to all the neutralino
  masses against $M_1$. Frame (a) is for scheme 1a, and frame (b) is
  for scheme 1b. Note that the two frames employ different scales
  along the $y-$axis.} 
\label{comps1s2}
\end{figure}

In Figure~\ref{fs1mu} we plot the corrections to the neutralino
masses in {\bf scheme 1} against $\mu$, with the left (right) frame
corresponding to scheme 1a (1b). Both frames show singularities at
$\mu = M_2$. The reason is that, in these schemes, the expressions for
$\delta M_2$ and $\delta \mu$ are undefined at $M_2 = |\mu|$, because
$\delta M_2$ and $\delta \mu$ are proportional to $\left( M_2^2 -
  \mu^2 \right)^{-1}$ \cite{Guasch, Baro2}. Hence the renormalized
masses are undefined if $M_2 = |\mu|$.  Moreover, some corrections
remain relatively large as long as $M_2$ and $|\mu|$ are close to each
other. In particular, the mass of $\tilde\chi_3^0$, which is mostly
higgsino--like in this region, receives a large correction, mostly
from $\delta \mu$. This pole is rather broad, e.g.  $\Delta^{0}_{3}$
can exceed $20\%$ if $\left| |\mu| - M_2 \right| \leq 5$ GeV.

Figs.~\ref{fs1mu} also show discontinuities at $\mu \simeq 131.50$
GeV. Here a level crossing occurs, i.e. the absolute value of the
negative eigenvalue of the neutralino mass matrix $M^{\rm n}$, which
used to give $m_{\tilde\chi_3^0}$, decreases past a positive
eigenvalue of $M^{\rm n}$ to give $m_{\tilde\chi_2^0}$. However,
if we track the corrections corresponding to the negative eigenvalue,
there is no discontinuity. A similar situation is encountered for $\mu
\sim -100 $ GeV, which is not shown in our plots. 

With this set of parameters, schemes 1a and 1b perform equally well,
i.e. the corrections have similar magnitudes. The results are
different for $|\mu| \leq M_1$, when the lightest neutralino is no
longer bino--like. This leads to another discontinuity in scheme 1b at
$\mu \simeq 110$ GeV: for smaller (larger) $\mu$ the most bino--like
neutralino is $\tilde \chi_3^0 \ (\tilde \chi_1^0)$. When approaching
this point from above, $\Delta^0_3$ therefore switches from a positive
value to zero, while $\Delta^0_1$ becomes non--vanishing (and
negative, for this choice of parameters). Moreover, $\delta M_1$
becomes large in scheme 1a if $|\mu| \ll M_1$, as anticipated in the
discussion in Subsec.\ref{sec:schemes}; see also
Table~\ref{schemes}. This leads to a sizable correction to the mass of
the bino--like neutralino.

This does not cause serious instability in Fig.~\ref{fs1mu}, since the
small value of $M_1$ chosen implies that $\left| M_1 - \mu \right|$
cannot exceed $M_Z$, i.e. $\tilde \chi_1^0$ retains a significant bino
component for all values of $\mu$ shown. Note also that LEP searches
for $\tilde \chi_1^+ \tilde \chi_1^-$ and $\tilde \chi_1^0 \tilde
\chi_2^0$ production anyway exclude the region $\mu \leq 120$ GeV.
The problematic behavior of scheme 1a for $|\mu| < M_1 - M_Z$ becomes
more evident in Fig.~\ref{comps1s2}, where we vary $M_1$ instead of
$\mu$. As $M_1$ grows large, the bino component of the lightest
neutralino decreases, $N_{11} \propto M_Z / (M_1 - \mu)$. Hence
$\delta M_1$ becomes large in scheme 1a, giving a large correction to
the mass of the bino--like neutralino, which is $\tilde \chi_4^0$ for
$M_1 > M_2$ in this figure. Note that in anomaly--mediated
supersymmetry breaking one expects $M_1 \simeq 2.8 M_2$
\cite{Drees2}. As a consequence $|M_1| - |\mu| > M_Z$ can easily be
realized in such a scenario.

Fig.~\ref{comps1s2} shows some discontinuities at $M_1 \simeq 335$
GeV. Just below this value $\tilde \chi_2^0$ is the most bino--like
neutralino, but at $M_1 = 335$ GeV $\tilde \chi_4^0$ becomes more
bino--like. Shortly thereafter, at $M_1 \simeq 340$ GeV, there is a
level crossing between $\tilde \chi_2^0$ and $\tilde \chi_3^0$; the
corresponding eigenvalues of the neutralino mass matrix have opposite
sign. Note that all corrections remain below 1\% in magnitude near
these discontinuities.

\begin{figure*}[t]
\epsfig{file=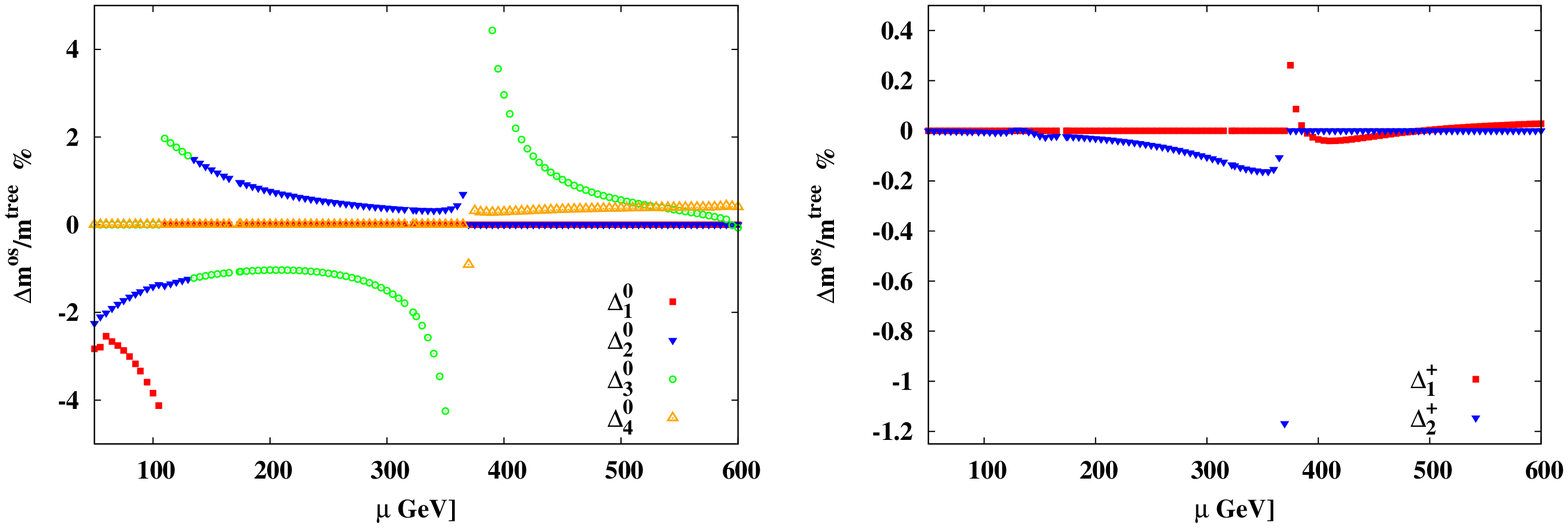, width=\linewidth, height=6 cm}
$~~~~~~~~~~~~~~~~~~~~~~~~~~~~~~ \text{(a)}  \hspace{7.75cm} \text{(b)}$
\caption{Radiative corrections (in \%) to (a) the neutralino masses
  and (b) the chargino masses against $\mu$. Results are for scheme
  2b, where the masses of the higgsino--like chargino and of the
  wino-- and bino--like neutralinos are used as inputs.}
\label{fs3mu}
\end{figure*}

\begin{figure*}[!ht]
\epsfig{file=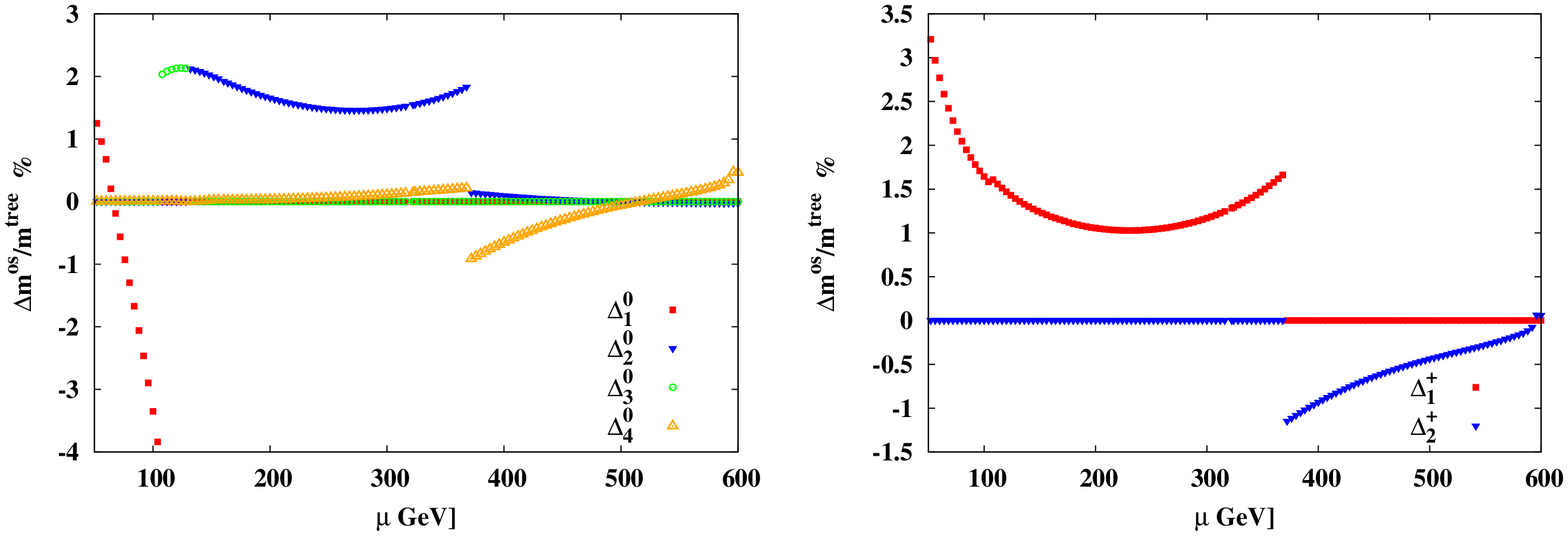, width=\linewidth, height=6 cm}
$~~~~~~~~~~~~~~~~~~~~~~~~~~~~~~ \text{(a)}  \hspace{7.75cm} \text{(b)}$
\caption{Radiative corrections (in \%) to (a) the neutralino masses
  and (b) the chargino masses against $\mu$. Results are for scheme
  2c, where the masses of the wino--like chargino and of the bino--
  and higgsino--like neutralinos are used as inputs.}
\label{fs4mu}
\end{figure*}
\begin{figure*}[!ht]
\epsfig{file=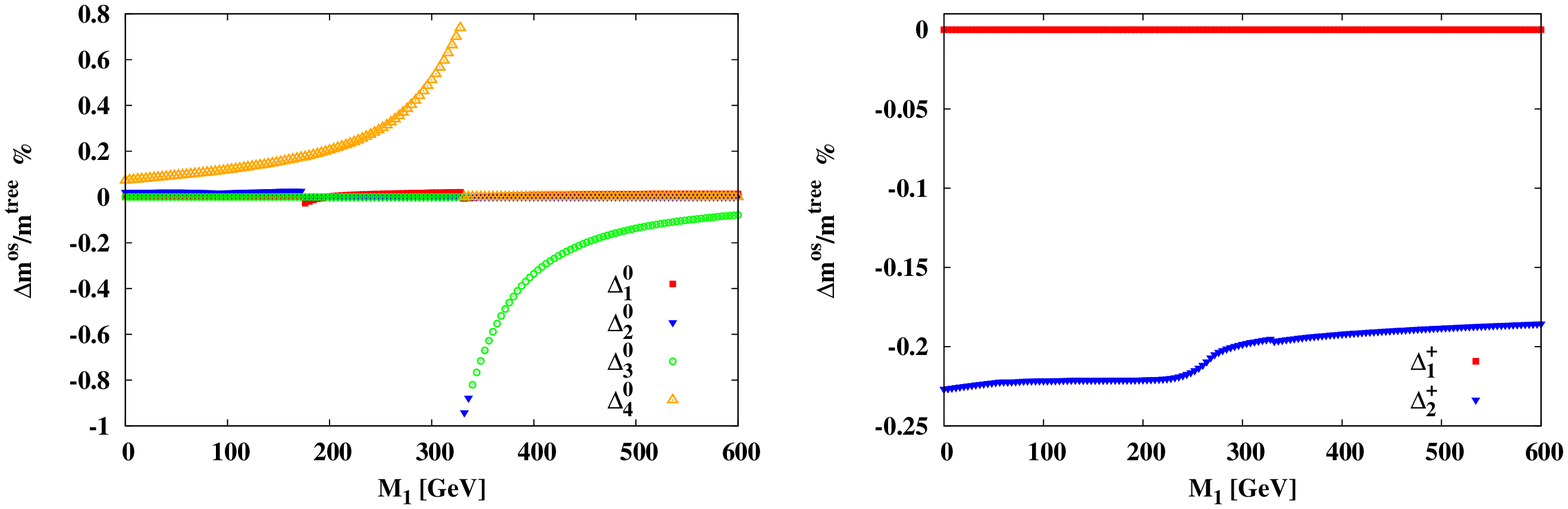, width=\linewidth, height=6 cm}
$~~~~~~~~~~~~~~~~~~~~~~~~~~~~~~ \text{(a)}  \hspace{7.75cm} \text{(b)}$
\caption{Radiative corrections (in \%) to (a) the neutralino masses
  and (b) the chargino masses against $M_1$. Results are for scheme
  2c, where the masses of the wino--like chargino, bino-- and higgsino--like
  neutralinos are used as inputs.}
\label{fs4m1}
\end{figure*}

\begin{figure*}[!ht]
\epsfig{file=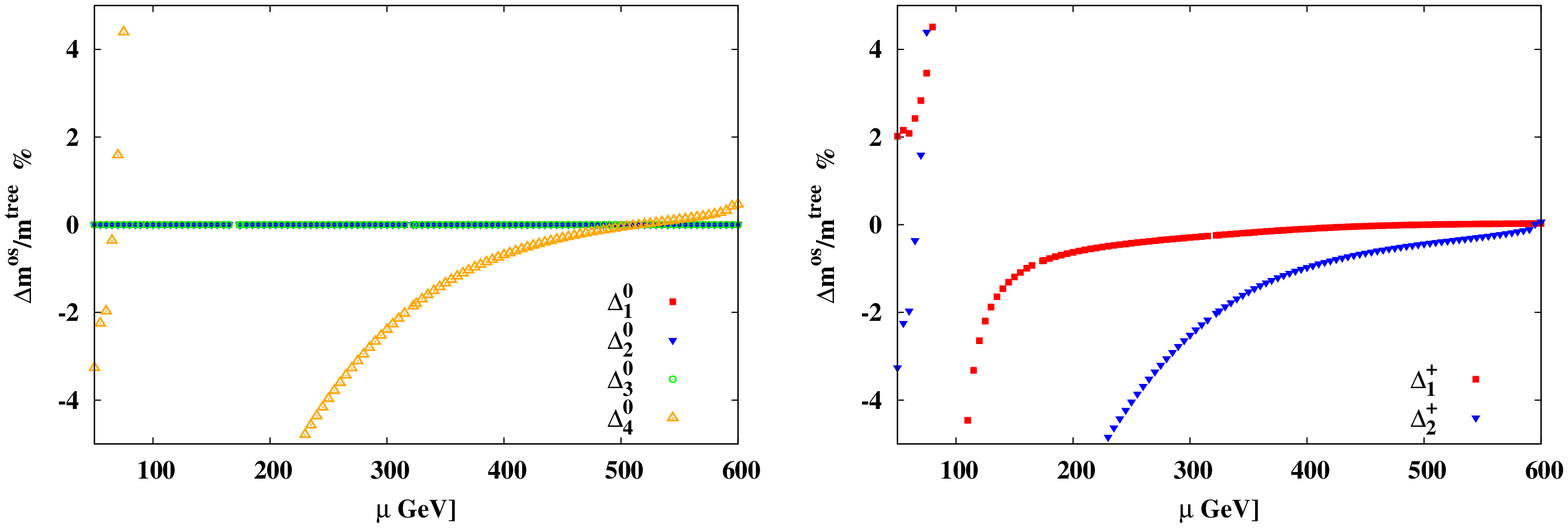, width=\linewidth, height=6 cm}
$~~~~~~~~~~~~~~~~~~~~~~~~~~~~~~ \text{(a)}  \hspace{7.75cm} \text{(b)}$
\caption{Radiative corrections (in \%) to (a) the neutralino masses
  and (b) the chargino masses against $\mu$. Results are for scheme
  3a, where the masses of the lightest three neutralinos are used as
  inputs.}
\label{fs5mu}
\end{figure*}
\begin{figure*}[!ht]
\epsfig{file=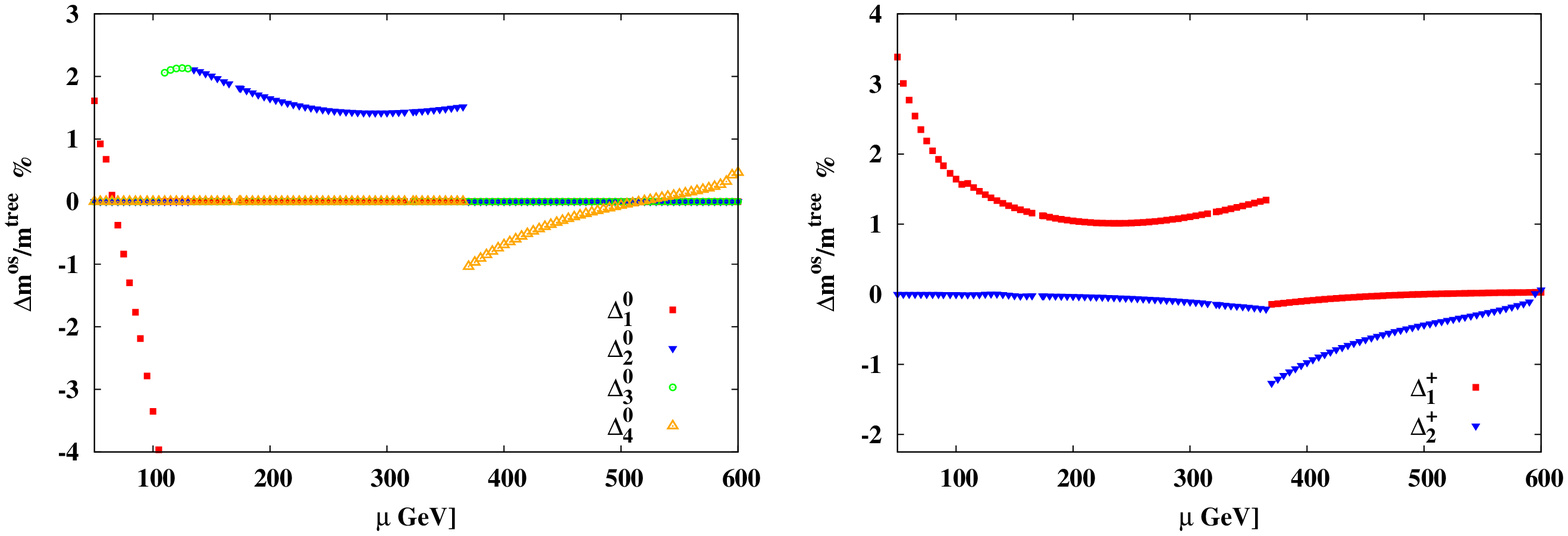, width=\linewidth, height=6 cm}
$~~~~~~~~~~~~~~~~~~~~~~~~~~~~~~ \text{(a)}  \hspace{7.75cm} \text{(b)}$
\caption{Radiative corrections (in \%) to (a) the neutralino masses
  and (b) the chargino masses against $\mu$. Results are for scheme
  3b, where the masses of the bino--, wino-- and higgsino--like
  neutralinos are used as inputs.}
\label{fs6mu}
\end{figure*}
\begin{figure*}[!ht]
\epsfig{file=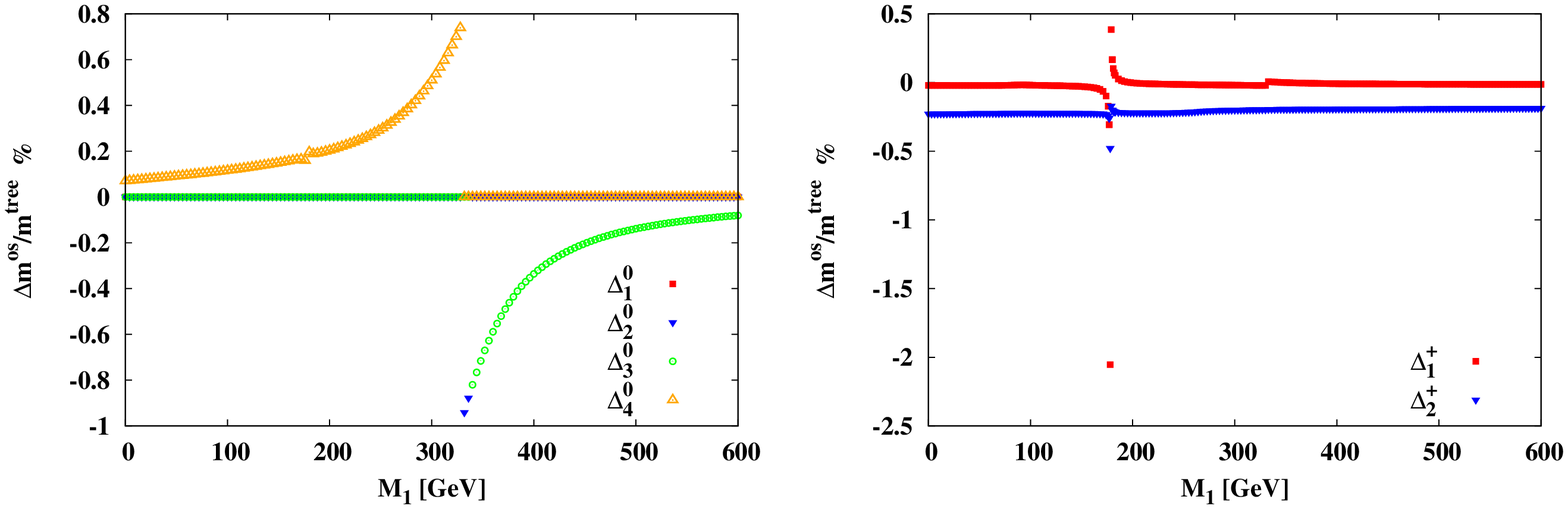, width=\linewidth, height=6 cm}
$~~~~~~~~~~~~~~~~~~~~~~~~~~~~~~ \text{(a)}  \hspace{7.75cm} \text{(b)}$
\caption{Radiative corrections (in \%) to (a) the neutralino masses
  and (b) the chargino masses against $M_1$. Results are for scheme
  3b, where the masses of the bino--, wino-- and higgsino--like
  neutralinos are used as inputs.}
\label{fs6m1}
\end{figure*}


In Fig.~\ref{fs3mu}, we demonstrate that scheme 2b also performs
equally well with our choice of parameters. However, the counterterms
again show a singularity at $M_2 \simeq |\mu|$ (although not exactly
at $M_2 = |\mu|$), where the denominator $D$ of the counterterms,
given in Eq.(\ref{D2}), vanishes. This again leads to large
corrections to the mass of the higgsino--like neutralino $\tilde
\chi_3^0$ in the vicinity of the singularity, very similar to both
versions of scheme 1. 

In contrast, we find that scheme 2a (not shown) performs poorly
whenever $M_2$ and $|\mu|$ are very different, as anticipated in the
discussion of Table~\ref{schemes}. In fact, numerically it is
considerably worse than scheme 1a. In the latter only $\delta M_1$ is
occasionally poorly determined; the corresponding corrections are
determined by the $U(1)_Y$ coupling. In scheme 2a, $\delta \mu$ or
$\delta M_2$ can be poorly determined, which receive corrections
proportional to the larger $SU(2)$ coupling. As a result the region of
parameter space with large corrections is not only larger in scheme 2a
than in scheme 1a, the size of the correction in these regions also
tends to be larger.

Scheme 2c does not have any singularity near $M_2 = |\mu|$. In
Fig.~\ref{fs4mu} it produces stable and small corrections to all three
chargino and neutralino masses that are not inputs for the entire
range of $\mu$. The curves in this figure show several
discontinuities, but these are harmless, since they do not endanger
perturbative stability. In particular, in frame (a) we observe the
same discontinuity as in Fig.~\ref{fs1mu}, since the states
$\tilde\chi_3^0$ and $\tilde\chi_2^0$ cross each other at
this point. The discontinuities at $\mu \simeq 110$ GeV have the same
origin as in scheme 1b, see the discussion of 
Fig.~\ref{fs1mu}(b). Finally, at $\mu \simeq 370$ GeV (i.e. $M_2
\simeq \mu$), the wino--like state changes from $\tilde\chi_2^+$ to
$\tilde\chi_1^+$ as $\mu$ increases. This affects $\delta M_2$
significantly, giving rise to (harmless) discontinuities in both
frames of Fig.~\ref{fs4mu}. 

In Fig.~\ref{fs4m1} we illustrate results for scheme 2c as function of
$M_1$. The discontinuities in the corrections to neutralino masses
near $M_1 = 335$ GeV have the same origin as in scheme 1b, see
Fig.~\ref{comps1s2}(b). All corrections are well behaved for
$|M_1| \simeq |\mu|$.

Fig.~\ref{fs5mu} demonstrates that scheme 3a also behaves poorly.  For
$\mu \lsim 200$ GeV all three input states, which are the lightest
three neutralinos in this scheme, have small wino component.  As a
consequence the $\delta M_2$ is poorly determined, leading to huge
corrections to the masses of the wino--like states $\tilde\chi_4^0$
and $\tilde\chi_2^+$, largely from $\delta M_2$. The scheme also shows
instability around $M_1 = \mu$, giving very large corrections to the
masses of both charginos and the wino--like neutralino.

Scheme 3b does not have any singularity near $M_2 = |\mu|$. In
Fig.~\ref{fs6mu}, like scheme 2c, it also produces stable and small
corrections to all three chargino and neutralino masses that are not
inputs for the entire range of $\mu$. In frame (a) we observe the same
discontinuity as in Fig.~\ref{fs1mu}, since the states
$\tilde\chi_3^0$ and $\tilde\chi_2^0$ cross each other at this
point. The discontinuities at $\mu \simeq 110$ GeV have the same
origin as in scheme 1b, see the discussion of Fig.~\ref{fs1mu}(b). The
harmless discontinuities at $\mu \simeq 370$ GeV (i.e. $M_2 \simeq
\mu$), are essentially the same as in Fig.~\ref{fs4mu}.

In Fig.~\ref{fs6m1} we show results for scheme 3b as function of
$M_1$. The discontinuities in the corrections to neutralino masses
near $M_1 = 335$ GeV have the same origin as in scheme 1b, see
Fig.~\ref{comps1s2}(b). In the case at hand this also leads to a small
discontinuity in $\Delta_1^+$. All corrections are well behaved for
$|M_1| \simeq |\mu|$, but there is some instability around $M_1 =
178.14$ GeV. Here the denominator $D$, given in Eq.(\ref{D3}), of the
counterterms vanishes. As a result, the neutralino and chargino masses
become ill--defined at this point. The effect of this pole is most
visible in the mass of the lightest chargino which, being wino--like,
receives dominant correction from $\delta M_2$. However, even this
corrections drops to $\sim 2\%$ when $M_1$ is just $1$ GeV away from
the pole, i.e. this pole is much narrower than the poles at $M_2 =
|\mu|$ in schemes 1a and 1b, and the poles at $M_2 \simeq |\mu|$ in
schemes 2a and 2b. These poles are discussed in more detail in the
following Subsection.

\subsection{Singularities in the Counterterms}
\begin{figure*}[!ht]
\epsfig{file=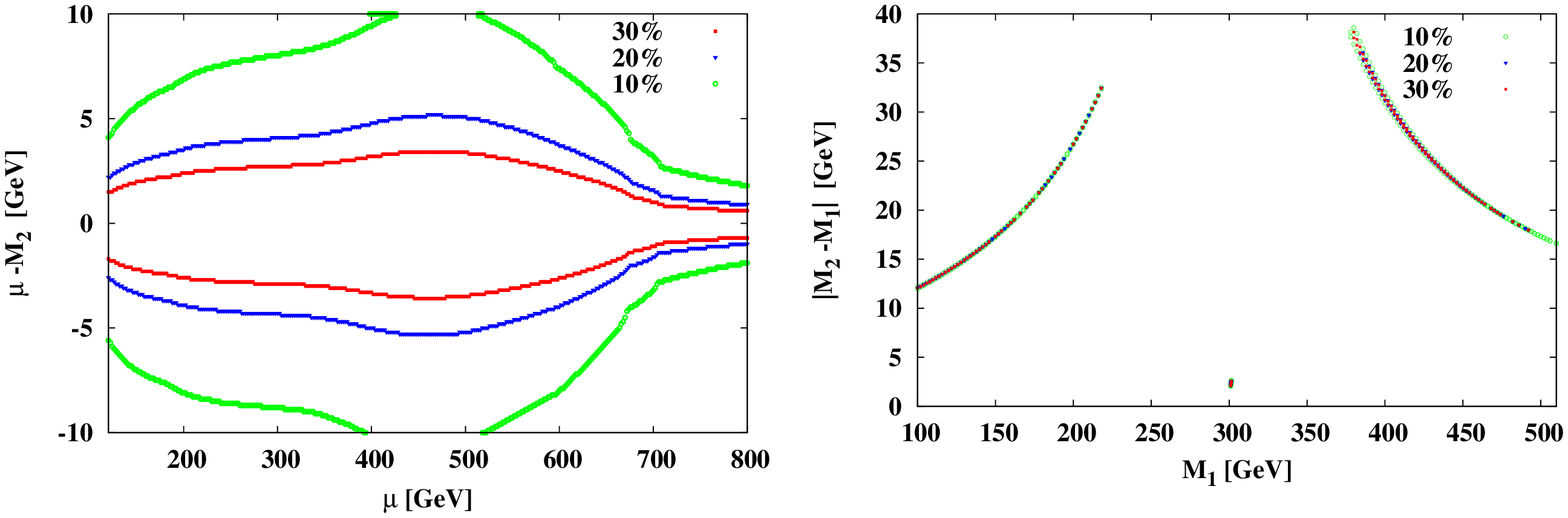, width=\linewidth, height=6 cm}
$~~~~~~~~~~~~~~~~~~~~~~~~~~~~~~ \text{(a)}  \hspace{7.75cm} \text{(b)}$
\caption{Contours of constant radiative corrections to (a) the
  higgsino--like neutralino mass around $M_2 \simeq \mu$ in scheme 1b,
  (b) the wino--like chargino mass around $M_1 \sim M_2$ in scheme 3b
  have been plotted against $\mu$ and $M_1$ respectively. Only
  corrections exceeding $10\%$ have been shown. The patch in (b), at
  $M_1 \sim 300 $ GeV, is actually at $M_1 \sim -300$ GeV, i.e. it is
  plotted against $-M_1$.}
\label{fig:poles}
\end{figure*}

In Fig.~\ref{fig:poles} the occurrence of poles, i.e. true divergences
in the ``finite'' parts of the counterterms (as opposed to harmless
discontinuities due to level crossings and the like), is analyzed
further. These are not due to divergent loop integrals, but due to
zeroes of the denominators in the expressions for the counterterms
[Eqs.(\ref{D2}) and (\ref{D3}), and the corresponding expression for
scheme 1]. We have assumed $M_1 = 180$ GeV and $\mu = 300$ GeV in
Figs.~\ref{fig:poles}a and \ref{fig:poles}b, respectively. Here we
show the maximal (relative) corrections. In scheme 1b these occur for
the higgsino--like neutralino, i.e. Fig.~\ref{fig:poles}a shows
$\Delta_2^0$ (for $\mu \lsim 168$ GeV as well as $168 \ {\rm GeV} \lsim
\mu \lsim 174 \ {\rm GeV}$ and $M_2 > \mu$) or $\Delta_3^0$ (for $\mu
\gsim 174$ GeV as well as $168 \ {\rm GeV} \lsim \mu \lsim 174 \ {\rm
  GeV}$ and $M_2 < \mu$).\footnote{In the region $168 \ {\rm GeV} \leq
  \mu \leq 174 \ {\rm GeV}$ the higgsino--like neutralino trades
  places with a photino--like state. Since these states have
  eigenvalues of the neutralino mass matrix with different sign, this
  cross--over of states is not associated with strong additional
  mixing, and thus does not cause any feature in
  Fig.~\ref{fig:poles}a.} In scheme 3b the mass of the wino--like
chargino instead receives the largest correction,
i.e. Fig.~\ref{fig:poles}b, shows $\Delta_2^+$ (for $M_2 > \mu$) or
$\Delta^{+}_{1}$ (for $M_2 < \mu$). Finally, scheme 2b shows
instability {\em both} at $M_2 \simeq |\mu|$ {\em and} at $M_1 \simeq
M_2$.

As already mentioned, it is easy to see that $\delta M_2$ and $\delta
\mu$ diverge in both versions of scheme 1 at $M_2 = |\mu|$
\cite{Guasch,Baro2}. In schemes 2 and 3 the expressions for the
denominators of the counterterms, given in Eqs.(\ref{D2}) and
(\ref{D3}), respectively, are too complicated to allow the derivation
of analytical expressions for the pole positions in terms of the input
parameters $M_1, \, M_2, \, \mu$ and $\tan\beta$. However, it is easy
to see that in the limit of small mixing only one term in each of
these expressions is large, leading to non--vanishing (and sizable)
denominators. Cancellations leading to poles can only occur in the
presence of large mixing, which requires at least one of the
differences $\left| |M_1| - M_2 \right|, \ \left| |M_1| - |\mu|
\right|,\ \left| M_2 - |\mu| \right|$ to be relatively small. Note that
this is a necessary condition, not a sufficient one.

In fact, Eqs.(\ref{D2}) and (\ref{D3}) show that singularities can
only occur if at least {\em two} of the input states are strongly
mixed. This explains the absence of poles at $M_2 \simeq |\mu|$ in
schemes 2c and 3b. Here the most higgsino--like neutralino is chosen
as input state. Note that even at $M_2 = |\mu|$ only one of the two
higgsino--like neutralinos mixes strongly with the wino; the other
state corresponds to an eigenvalue with opposite sign ($\simeq -|\mu|$),
and shows little mixing. Small mixing means large higgsino component,
i.e. this weakly mixed state is picked up by our algorithm as the most
higgsino--like one. At $M_2 = |\mu|$ (and sufficiently large $\left|
  |M_1| - M_2 \right|$) the input states in these schemes are thus a
relatively pure bino (neutralino), a relatively pure higgsino
(neutralino), and a strongly mixed wino--higgsino state (a chargino in
scheme 2c, a neutralino in scheme 3b). The fact that two input states
show little mixing is sufficient to single out one term in
eqs.(\ref{D2}) and (\ref{D3}) as dominant one, making significant
cancellations impossible. A similar argument explains why none of our
schemes has instabilities for $|M_1| \simeq |\mu|$ (unless one {\em
  also} has $M_2 \simeq |\mu|$; see below).

Scheme 3b shows a singularity at $M_1 \simeq M_2$ where two input
neutralino states are strong mixtures of the bino and wino current
states, leading to a cancellation between the last two terms in
Eq.(\ref{D3}) with all other terms being small. Note that the pole
occurs only when $|M_1 - |\mu|| > |M_2 -|\mu||$.\footnote{For exact
  equality, $M_1 = M_2$, one eigenstate is an exact photino with mass
  $M_2$, whose bino--component is given by $\cos \theta_W \simeq
  0.88$, i.e. with relatively weak bino--wino mixing.} We did not find
any pole if $|M_1 - \mu| \lesssim 80$ GeV. In this interval one of the
bino--wino mixed states (usually the more wino--like one) also has a
substantial higgsino component, and correspondingly reduced gaugino
components. Thus there are no longer two input states with very
similar bino and wino components.

This pole at $M_1 \simeq M_2$ is very narrow. This is partly due to
cancellations of the numerators of the expressions for the
counterterms, which can be understood as follows. Near the pole,
indices $i$ and $j$ denote strong mixtures between bino and wino
states, and $k$ denotes a higgsino state with little mixing (unless
$\left| |\mu| - M_2 \right|$ is also small). In that case only two
terms in each of Eqs.(\ref{dmu3})--(\ref{dm13}) are not
suppressed. These terms are antisymmetric in $i \leftrightarrow j$,
hence strong mixing also induces strong cancellations; however, these
cancellations are not perfect, since mixing pushes the masses
$m_{\tilde \chi_i^0}$ and $m_{\tilde \chi_j^0}$ apart. 

The finite parts of the counterterms themselves can therefore become
quite large, often exceeding the tree--level values of these
parameters. Even then corrections to physical masses need not be
large, due to cancellations in Eqs.(\ref{m_pchar}) and
(\ref{m_pneut}). By construction these cancellations are perfect for
the three input states. If an output state is of similar nature as one
of the input states, there will then also be strong cancellations for
the mass of this output state. For example, in scheme 1b with $M_2
\simeq |\mu|$, the strong wino--higgsino mixtures in the neutralino
sector have similar wino and higgsino components as the two input
chargino states, leading to strong cancellations. The largest
correction then results for the (almost) unmixed higgsino state in the
neutralino sector, as shown in Fig.~\ref{fig:poles}a. The same
argument explains why in the ``generic'' situation with relatively
small mixing, in all our b and c schemes the mass of the most strongly
mixed state will receive the largest correction, since it is most
different from all the input states, which are the least mixed ones in
these schemes.

In scheme 3b with $M_1 \simeq M_2$, both strongly mixed gaugino--like
neutralinos, which are the cause of the pole, are input states, and
hence have perfect cancellations. These states mix relatively weakly
with the higgsino--like states. The counterterm $\delta \mu$, which
diverges at the pole, comes with coefficient $- 2 N_{k3}^* N_{k4}^*$
in Eq.(\ref{m_pneut}). In the limit of vanishing higgsino--gaugino
mixing, this is the same for the two higgsino--like states, one having
$N_{k3} \simeq N_{k4} \simeq i/\sqrt{2}$ (purely imaginary), while the
other had $N_{k3} \simeq - N_{k4} = 1/\sqrt{2}$ (purely real, but with
opposite signs). This explains why the correction to the mass of the
single output state in the neutralino sector remains very small near
this pole, see Fig.~\ref{fs6m1}.

One of the two chargino states is also higgsino--like, and the
cancellation will go through. However, the second chargino is
wino--like. Since there is no wino--like neutralino, only strongly
mixed ones, the cancellation in the equation for the mass of this
wino--like chargino is quite incomplete. This explains why this state
receives the largest correction, as shown in Fig.~\ref{fig:poles}b.

The cancellations we discussed so far work (to some extent) near all
poles, and help to confine their effects to quite narrow strips of
parameter space, as shown in Fig.~\ref{fig:poles}. The fact that the
instability region of scheme 3b at $M_1 \simeq M_2$ is much narrower
than that of scheme 1b (or 2b) at $M_2 \simeq |\mu|$ is due to the
fact that wino--bino mixing, which causes the former, only occurs at
second order in an expansion of the neutralino masses and eigenstates
in powers of $M_Z$, while wino--higgsino mixing, which causes the
latter, already occurs at first order; wino--bino mixing is also
suppressed by an extra factor of $\sin \theta_W$. As a result,
significant wino--bino mixing occurs over a much smaller region of
parameter space than large wino--higgsino mixing.

Finally, if $-M_1 \simeq M_2 \simeq |\mu|$ there are two
bino--higgsino mixed states and two wino--higgsino mixed states in the
neutralino sector. Thus scheme 3b also suffers from instability around
this point because two input states are strongly mixed.\footnote{For
  tan $\beta =1$ there is a pure higgsino state in the neutralino
  sector. The corresponding eigenvalue is given by $-\mu$. Thus, even
  with $ -M_1 \simeq M_2 \simeq |\mu|$, as tan $\beta$ approaches 1,
  there is only little bino--higgsino (wino--higgsino) mixing if $\mu
  > 0$ ($\mu < 0$).}  This explains the patch in Fig.~\ref{fig:poles}b
at $M_1 \sim -300$ GeV. Note that here the ``finite'' parts of the
counterterms diverge only in a very small region in the $(M_1, M_2)$
plane, extending for about 1 MeV in $M_2$ and about 300 MeV in $M_1$;
this instability is thus qualitatively different from the one at $M_1
\simeq M_2$, where a pole occurs along a line that is unbounded
towards large $M_2$.\footnote{In this patch the two neutralino states
  that remain after identifying the most bino-- and wino--like
  neutralinos have very similar higgsino components. The instability
  can be avoided by simply using the {\em other} higgsino--like state
  as input state, even though by our definition it is slightly less
  higgsino--like. Note also that disentangling the states
  experimentally will be very difficult in this region, again due to
  the strong mixing.}

The singularity at $M_1 \simeq M_2$ does not exist in scheme 2c,
since in this scheme again only one of the three input states, the
most bino--like neutralino, is strongly mixed in this region of
parameter space. The input chargino state is mostly wino--like, unless
one {\em in addition} has $M_2 \simeq |\mu|$, and the second input
neutralino is again mostly higgsino (even if $M_2 \simeq |\mu|$, as
explained above). 

For fixed $\mu$ and $\tan\beta$ Scheme 2c can therefore at worst
become singular in a very narrow patch in the $(M_1, M_2)$ plane, in
the region $-M_1 \simeq M_2 \simeq |\mu|$. We saw above that this
leads to two bino--higgsino mixed states and two wino--higgsino mixed
states in the neutralino sector. The chargino sector also contains two
strongly mixed (wino--higgsino) states. Note that the denominator $D$
of Eq.(\ref{D2}) is not a continuous function of the input parameters,
since the indices $i,j,k$ defining the input states take different
values in different regions of parameter space. This makes it rather
difficult to systematically search for zeroes of this denominator. A
straightforward numerical scan of $M_1$ and $M_2$ with step size of
$0.01$ GeV did not find any evidence for a zero of the denominator.
We did find points with large (finite parts of) $\delta M_1,\, \delta
M_2$ and $\delta \mu$ in a very narrow region of parameter space,
similar to the patch in Fig.~\ref{fig:poles}b. However, even in that
region the renormalized masses remain well behaved in scheme 2c, the
maximal correction not exceeding $4\%$. This is due to the two types
of cancellation discussed above. Note that all our other schemes lead
to considerable larger corrections around this point. We thus conclude
that scheme 2c is the most stable of the seven schemes we
investigated.

\subsection{Comparison Between Schemes}

It is generally expected that the physical masses calculated using two
different renormalization schemes at one--loop level can differ only by
terms of two--loop order. In the case at hand this is true whenever the
schemes are well behaved. We demonstrate this explicitly by comparing
results for schemes 1b and 3b. We then show another example where this
no longer holds, since one of the schemes produces large corrections
close to one of the singularities discussed in the previous
Subsection. We do not discuss schemes 1a, 2a and 3a in this
Subsection, since we already saw that they perform poorly in large
regions of parameter space, see Table~\ref{schemes}.

We use the following ``tree level" parameters to calculate the
on--shell renormalized masses in scheme (1b):
$$\text{Set~A:}~~M_1 = 90 ~\text{GeV},~ M_2 = 200 ~\text{GeV},~ \mu = 300
~\text{GeV},~ \tan \beta =5. $$ 
The masses thus obtained are shown in column 3 of Table~\ref{comp131}.
We then take $m_{\tilde\chi_1^0}$, $m_{\tilde\chi_2^0}$ and 
$m_{\tilde\chi_3^0}$, thus obtained, as input masses for scheme 3b. 
Note that these are respectively the masses of the most bino--, wino--
and higgsino--like neutralino. From these three masses, we compute
numerically $M_1$, $M_2$ and $\mu$:
$$\text{Set~B:}~~M_1 = 89.9697 ~\text{GeV},~ M_2 = 199.7935 ~\text{GeV},~ \mu =
301.0052 ~\text{GeV},~ \tan \beta =5.$$ We then calculate the
one--loop masses in scheme 3b. The resulting spectrum is shown in
column 4 of Table~\ref{comp131}. We also list in parentheses the
one--loop masses calculated using parameter set A within scheme 3b. As
shown in column 5, the differences between the two one--loop
predictions, using scheme 1b with Set A and scheme 3b with set B, are
much smaller than the differences between tree--level and one--loop
corrections (except for the input masses, of course, where the latter
differences vanish). This is a non--trivial check of our calculation.

Note that using the same numerical values for $M_1, \, M_2$ and $\mu$
in both schemes leads to differences between one--loop masses that are
themselves of one--loop order, as shown by the numbers in parentheses
in column 5 of Table~\ref{comp131}. Only after fixing three {\em
  physical} masses to be the same in both schemes do the predictions
for all other masses become nearly identical. The reason is that $M_1,
\, M_2$ and $\mu$ are themselves scheme dependent
quantities.

\begin{table}[h]  
\begin{center}
\begin{tabular}{|| c | c | c | c | c || }
\hline 
Particle & $m^{\rm tree}_{\tilde \chi} \ [{\rm GeV}]$ & 
\multicolumn{2}{c|}{$m^{\rm 1-loop, \, os}_{\tilde \chi} \ [{\rm
    GeV}]$} & $\delta m_{\tilde \chi} \ [{\rm GeV}]$  \\
        & [set A] & scheme 1b & scheme 3b  &  col. (3) -col. (4)\\
                        &      &  [set A] & [set B (set A)] & \\
\hline 
$\tilde\chi_1^+$ & $170.4007$ & $170.4007$ & $170.4030~ (170.3714)$ &
$-0.0023~ (0.0293)$ \\ 
$\tilde\chi_2^+$ & $337.5026$ & $337.5026$ & $337.5169~ (336.7472)$ &
$-0.0143~ (0.7554)$ \\ 
$\tilde\chi_1^0$ & $84.8425$ & $84.8425$ & $84.8425~ (84.8425)$ & $
0.0~ (0.0)$ \\
$\tilde\chi_2^0$ & $172.1478$ & $172.1545$ & $172.1545~ (172.1478)$ &
$0.0~ (0.0067)$ \\
$\tilde\chi_3^0$ & $305.4830$ & $306.4778$ & $306.4778~ (305.4830)$ &
$0.0~ (0.9948)$ \\
$\tilde\chi_4^0$ & $338.4927$ & $339.6138$ & $339.6313~ (338.8642)$ &
$-0.0175~ (0.7496)$ \\
\hline
\end{tabular}
\end{center}
\caption{Masses of the charginos and neutralinos. The second
  column lists tree--level predictions for input set A. Columns 3 and
  4 list one--loop predictions for schemes 1b and 3b; in the latter
  case, the first entry is for input set B, which has been chosen such
  that $m_{\tilde \chi_1^0},\, m_{\tilde\chi_2^0}$ and $m_{\tilde
    \chi_3^0}$ have the same values as in scheme 1b (third column),
  whereas the numbers in parentheses use input set A. The last column
  is the difference between the third and fourth columns.}

\label{comp131}
\end{table}  

However, as mentioned before, scheme 1b is undefined when $M_2 =
|\mu|$. We also saw that for $M_2 \simeq |\mu|$ the finite part of
the counterterm $\delta \mu$ remains large, so that the mass of a
higgsino--like neutralino receives a large correction. This pole is
absent in scheme 3b. As a consequence, for the same set of physical
masses, the one--loop masses in these two schemes can be quite
different. In Table~\ref{comp132} we demonstrate this by explicit
computation.

We use the following tree level parameters to calculate the on--shell 
renormalized masses in scheme 1b:  
$$\text{Set~C:}~~  M_1 = 90 ~\text{GeV}, M_2 = 350 ~\text{GeV}, \mu =
355 ~\text{GeV}, \tan \beta =5. $$ The resulting physical masses are
given in column 3 of table (\ref{comp132}). Note the very large
correction ($> 50$ GeV) to $m_{\tilde \chi_3^0}$.\footnote{In case of
  $m_{\tilde \chi_2^0}$ and $m_{\tilde \chi_4^0}$ large finite
  corrections from $\delta M_2$ and $\delta \mu$ largely cancel,
  similar to the exact cancellation required to keep both chargino
  masses unchanged. $\tilde \chi_3^0$ is a nearly pure higgsino; its
  mass is thus not sensitive to $\delta M_2$, so that no cancellation
  can take place.}  As before, we then take $m_{\tilde\chi_1^0}$,
$m_{\tilde\chi_0^2}$ and $m_{\tilde\chi_3^0}$, thus obtained, as input
masses for scheme 3b.  These are respectively the masses of the
bino--, wino-- and higgsino--like neutralino. From these three masses,
we compute numerically $M_1$, $M_2$ and $\mu$:
$$\text{Set~D:}~~M_1 = 89.2056  ~\text{GeV},~ M_2 = 319.1479
~\text{GeV},~ \mu = 418.1529 ~\text{GeV},~ \tan \beta =5.$$ 
Note that $\mu$ had to be increased by more than 60 GeV in order to
reproduce the one--loop prediction for $m_{\tilde \chi_3^0}$ from
scheme 1b. This increase of $\mu$ would also have led to sizable
increases of $m_{\tilde \chi_1^+}$ and $m_{\tilde\chi_2^0}$, had we
not simultaneously reduced $M_2$ by more than 30 GeV. Only $M_1$
retains approximately the same value in both Sets.

\begin{table}[h]  
\begin{center}
\begin{tabular}{|| c | c | c | c | c || }
\hline 
Particle & $m^{\rm tree}_{\tilde \chi} \ [{\rm GeV}]$ & 
\multicolumn{2}{c|}{$m^{\rm 1-loop, \, os}_{\tilde \chi} \ [{\rm
    GeV}]$} & $\delta m_{\tilde \chi} \ [{\rm GeV}]$  \\
        & [set C] & scheme 1b & scheme 3b  &  col. (3) -col. (4)\\
                        &      &  [set C] & [set D (set C)] & \\
\hline 
$\tilde\chi_1^+$ & $288.3284$ & $288.3284$ & $288.0337~ (287.9178)$ &
$0.2947~ (0.4106)$ \\
$\tilde\chi_2^+$ & $422.2986$ & $422.2986$ & $452.4840~ (417.3805)$ & 
$-30.1854~ (4.9181)$ \\ 
$\tilde\chi_1^0$ & $86.3675$  & $86.3675$ & $86.3675~ (86.3675)$ & $
0.0~ (0.0)$\\
$\tilde\chi_2^0$ & $289.8939$ & $288.7260$ & $288.7260~ (289.8939)$ & 
$0.0~ (-1.1679)$ \\
$\tilde\chi_3^0$ & $359.1238$ & $421.9883$ & $421.9883~ (359.1238)$ &
$0.0~ (62.8645)$ \\
$\tilde\chi_4^0$ & $422.8623$ & $424.2374$ & $454.7079~ (418.8926)$ &
$-30.4705~ (5.3448)$ \\ 
\hline
\end{tabular}
\end{center}
\caption{Masses of neutralinos and charginos. The meaning of the
  various entries is as in Table~\ref{comp131}, except that parameter
  sets A and B have been replaced by sets C and D, respectively.}
\label{comp132}
\end{table} 

The resulting one--loop masses in scheme 3b are listed in column 4 of
Table~\ref{comp132}. In parentheses we also give the one--loop masses
calculated using parameter set C using scheme 3b. Column 5 shows that
the differences between the one--loop predictions for
$m_{\tilde\chi_2^+}$ and $m_{\tilde\chi_4^0}$ are very
large. This is due to the large differences in the input parameters
between Set C and Set D, which in turn was enforced by the large
correction to $m_{\tilde \chi_3^0}$ in scheme 1b. 

We emphasize that in the case at hand, the predictions of scheme 1b
are clearly not reliable, due to the vicinity of the pole at $M_2 =
\mu$ which leads to a breakdown of the perturbative expansion in this
scheme. In scheme 3b the differences between tree--level and one--loop
masses do not exceed 1\% for both set C and set D. Moreover,
differences between physical masses are again very small if scheme 3b
is compared to scheme 2c, which also has no pole at $M_2 \simeq \mu$.


\section{Conclusion}
\label{sec:con}

In this paper we have analyzed variants of the on--shell
renormalization of the chargino and neutralino sector of the
MSSM. There are six physical particles with in general different
masses in this sector, but only three free parameters whose
counterterms are determined from this sector. Hence only three masses
can be chosen as ``input masses'', which by definition are unchanged
to all orders in perturbation theory; the remaining three masses will
receive quantum corrections, which we computed to one--loop order.

We introduced seven different versions of the on--shell scheme in
Sec.~2. These schemes differ by the states whose masses are used as
inputs. General considerations led us to expect that all schemes where
the indices of the three input states are fixed a priori can show
instability over large regions of parameter space, as listed in
Table~\ref{schemes} for schemes 1a, 2a and 3a; we confirmed this by
explicit calculations in Subsec.~4.1. In these regions at least one of
three counterterms $\delta M_1, \, \delta M_2$ and $\delta \mu$ is
poorly determined, since none of the input masses is very sensitive to
it. These large regions of instability can be avoided by choosing the
masses one bino--like, one wino--like and one higgsino--like state
as input; the definition of these states, and other practical matters,
have been discussed in Sec.~3. This ensures that there is at least one
input mass that is sensitive to each counterterm.

Unfortunately this is not sufficient to avoid all regions of
instability. In particular, we see that all schemes where the mass of
the higgsino--like chargino is used as an input become unstable for
$M_2 \simeq |\mu|$, because finite parts of the counterterms $\delta
M_2$ and $\delta \mu$ become very large, leading to a very large
correction to the mass of the (most) higgsino--like neutralino. This
instability is caused by numerical cancellations in the denominators
of the expressions for the counterterms, which can only occur if at
least two input states are strongly mixed states. We saw in
Subsecs.~4.2 and 4.3 that this instability can lead to a collapse of
the perturbative expansion over a significant region of parameter
space. It can be avoided if instead the mass of the higgsino--like
neutralino is used as input; this state does not mix strongly even if
$M_2 \simeq |\mu|$ or $|M_1| \simeq |\mu|$.

Finally, we found another similar pole, with however in general much
smaller numerical values of the corrections somewhat away from the
pole, if the masses of both the bino-- and wino--like neutralino are
used as inputs. This instability is due to bino--wino mixing, which
occurs at $M_1 \simeq M_2$. It can be avoided by chosing the mass of
the wino--like chargino, rather than the wino--like neutralino, as
input, since the charginos are obviously not affected by bino--wino
mixing.

We conclude that the perturbatively most stable version of the
on--shell scheme is the one where the masses of the wino--like
chargino, and of the bino-- and higgsino--like neutralinos are used as
inputs; the scheme where the masses of the bino--, wino-- and
higgsino--like neutralinos are used as input performs as well over
most, but not all, of parameter space.

We emphasize that a poor choice of renormalization scheme can lead to
unphysically large corrections not only to some masses, but also to
cross sections. For example, the exchange of heavier neutralino states
can play an important role in the cross section for the annihilation
of a pair of the lightest neutralinos into final states containing
gauge and/or Higgs bosons \cite{Drees:1992am}; the total neutralino
annihilation cross section determines their present relic density if
they are stable. The proper choice of renormalization scheme should
therefore be important for a variety of one--loop calculations in the
MSSM that have been, and are being, performed.

\subsection*{Acknowledgment}
We would like to thank Thomas Hahn and Karina Williams for helping
with the installation of the Feyn packages, and Nicolas Bernal for
help with Gnuplot. This work was partially supported by the DFG
Transregio TR33 ``The Dark Side of the Universe'', and partly by the
German Bundesministerium f\"ur Bildung und Forschung (BMBF) under
Contract No. 05HT6PDA.

\section*{Appendix A}
\label{sec:appa}

\vspace{.5 cm}

Here we give expressions for $\delta M_1$, $\delta M_2$ and $\delta
\mu$ in terms of the $\delta m_{\tilde \chi}$ of Eq.(\ref{dmf}) as
well as the counterterms to the off--diagonal entries of the chargino
and neutralino mass matrices, which can be derived from
Eqs.(\ref{delta_SM}) and (\ref{delta_tanbeta}). These expressions have
been obtained by solving equations (\ref{scheme2ab}), i.e. they are
valid for all versions of {\bf scheme 2}, with different variants
corresponding to different choices of the indices $i,j,k$ (see
Table~\ref{schemes}). The counterterms can be written as
\bea
\delta M_{1} & = &\dfrac{N_{M_1}}{D},\\
\delta M_{2} & = &\dfrac{N_{M_2}}{D},\\
\delta \mu   & = & \dfrac{N_{\mu}}{2 D},
\eea
where
\beq \label{D2}
D =  2 N^*_{i3} N^*_{i4} N^{*^2}_{j1} U^*_{k1} V^*_{k1}-2
N^{*^2}_{i1} N^*_{j3} N^*_{j4} U^*_{k1} V^*_{k1}
+N^{*^2}_{i2} N^{*^2}_{j1} U^*_{k2} V^*_{k2}- N^{*^2}_{i1}
N^{*^2}_{j2} U^*_{k2} V_{k2}, 
\eeq
\beq
\beal
N_{\mu}  = -\left((\delta m_{\tilde\chi^0_i}-2 \delta M^n_{i3}
  N^*_{i1} N^*_{i3}- 2 \delta M^n_{23} N^*_{i2} N^*_{i3} -2
  \delta M^n_{14} N^*_{i1} N^*_{i4}  \right.\\  
 \left. - 2 \delta M^n_{24} N^*_{i2} N^*_{i4}) N^{*^2}_{j1} + N^{*^2}_{i1}
(-\delta m_{\tilde\chi^0_j}+ 2 \delta M^n_{13} N^*_{j1} N^*_{j3} 
+ 2 \delta M^n_{23} N^*_{j2} N^*_{j3} \right.\\ 
\left.  +2 \delta M^n_{14} N^*_{j1} N^*_{j4} + 2 \delta
  M^n_{24} N^*_{j2} N^*_{j4})\right)U^*_{k1} V^*_{k1}
+ \left(-N^{*^2}_{i2} N^{*^2}_{j1}+N^{*^2}_{i1} N^{*^2}_{j2}\right)\\ 
\left(-\delta m_{\tilde\chi^+_k}+\delta M^c_{21} U^*_{k2} V^*_{k1}+
\delta M^c_{12} U^*_{k1} V^*_{k2}\right),
\eeal
\eeq
\bea
N_{M_2} & = & 2 \delta m_{\tilde\chi^+_k} N^*_{i3} N^*_{i4}
N^{*^2}_{j1}-2 \delta m_{\tilde\chi^+_k} N^{*^2}_{i1} N^*_{j3}
N^*_{j4} -2 \delta M^c_{21} N^*_{i3}
N^*_{i4} N^{*^2}_{j1} U^*_{k2} V^*_{k1} \nonumber \\
& + & 2 \delta M^c_{21} N^{*^2}_{i1} N^*_{j3} N^*_{j4}
U^*_{k2} V^*_{k1}-2 \delta M^c_{12} N^*_{i3} N^*_{i4}
N^{*^2}_{j1} U^*_{k1} V^*_{k2}-\delta m_{\tilde\chi^0_j}
N^{*^2}_{i1} U^*_{k2} V^*_{k2}\nonumber \\ & + &
 \delta M^c_{12} N^{*^2}_{i1} N^*_{j3} N^*_{j4} U^*_{k1}
 V^*_{k2}+\delta m_{\tilde\chi^0_i} N^{*^2}_{j1} U^*_{k2}
 V^*_{k2} -2 \delta M^n_{13} N^*_{i1} N^*_{i3} N^{*^2}_{j1}
 U^*_{k2} V^*_{k2} \nonumber \\ 
& - & 2 \delta M^n_{23} N^*_{i2} N^*_{i3} N^{*^2}_{j1} U^*_{k2}
 V^*_{k2}-2 \delta M^n_{14} N^*_{i1} N^*_{i4} N^{*^2}_{j1} U^*_{k2}
 V^*_{k2}-2 \delta M^n_{24} N^*_{i2} N^*_{i4} N^{*^2}_{j1}
 U^*_{k2} V^*_{k2}\nonumber \\ 
& + &2 \delta M^n_{13} N^{*^2}_{i1} N^*_{j1} N^*_{j3}
 U^*_{k2} V^*_{k2}+2 \delta M^n_{23} N^{*^2}_{i1} N^*_{j2} N^*_{j3}
U^*_{k2} V^*_{k2}+2 \delta M^n_{14} N^{*^2}_{i1} N^*_{j1}
N^*_{j4} U^*_{k2} V^*_{k2}\nonumber \\ & + &2 \delta M^n_{24}
N^{*^2}_{i1} N^*_{j2} N^*_{j4} U^*_{k2} V^*_{k2},
\eea
\bea
N_{M_1}& = & 2 \delta m_{\tilde\chi^+_k} N^*_{i3} N^*_{i4}
N_{j2}^{*^2}-2 \delta m_{\tilde\chi^+_k} N_{i2}^{*^2} N^*_{j3}
N^*_{j4} -2 \delta m_{\tilde\chi^0_j} N^*_{i3} N^*_{i4}
U^*_{k1} V^*_{k1}-\delta m_{\tilde\chi^0_j} N_{i2}^{*^2}
U^*_{k2} V^*_{k2} 
\nonumber \\ 
& - & 2 \delta M^n_{14} N^*_{i1} N^*_{i4} N_{j2}^{*^2}
U^*_{k2} V^*_{k2}-2 \delta M^n_{24} N^*_{i2} N^*_{i4}
N_{j2}^{*^2} U^*_{k2} V^*_{k2} + 2 \delta M^n_{13} N_{i2}^{*^2}
N^*_{j1} N^*_{j3} U^*_{k2} V^*_{k2} 
\nonumber \\ 
& + & 2 \delta M^n_{23} N_{i2}^{*^2} N^*_{j2} N^*_{j3}
U^*_{k2} V^*_{k2}+2 \delta M^n_{14} N_{i2}^{*^2} N^*_{j1}
N^*_{j4} U^*_{k2} V^*_{k2} + 2 \delta M^n_{24} N_{i2}^{*^2}
N^*_{j2} N^*_{j4} U^*_{k2} V^*_{k2} 
\nonumber \\
&- & 2 \delta M^c_{12} N^*_{i3} N^*_{i4} N_{j2}^{*^2} U^*_{k1}
V^*_{k2}+2 \delta M^c_{12} N_{i2}^{*^2} N^*_{j3} N^*_{j4}
U^*_{k1} V^*_{k2} + 4 \delta M^n_{13} N^*_{i3} N^*_{i4}
N^*_{j1} N^*_{j3} U^*_{k1} V^*_{k1}
\nonumber \\ 
& + & 4 \delta M^n_{23} N^*_{i3} N^*_{i4} N^*_{j2} N^*_{j3}
U^*_{k1} V^*_{k1}-2 \delta M^n_{13} N^*_{i1} N^*_{i3}
N_{j2}^{*^2} U^*_{k2} V^*_{k2} - 2 \delta M^n_{23} N^*_{i2}
N^*_{i3} N_{j2}^{*^2} U^*_{k2} V^*_{k2} 
\nonumber \\
& + &
4 \delta M^n_{14} N^*_{i3} N^*_{i4} N^*_{j1} N^*_{j4}
U^*_{k1} V^*_{k1}+4 \delta M^n_{24} N^*_{i3} N^*_{i4}
N^*_{j2} N^*_{j4} U^*_{k1} V^*_{k1} + 2 \delta
m_{\tilde\chi^0_i} N^*_{j3} N^*_{j4} U^*_{k1} V^*_{k1} 
\nonumber \\
&- & 4 \delta M^n_{24} N^*_{i2} N^*_{i4} N^*_{j3} N^*_{j4}
U^*_{k1} V^*_{k1}-2 \delta M^c_{21} N^*_{i3} N^*_{i4}
N_{j2}^{*^2} U^*_{k2} V^*_{k1} + 2 \delta M^c_{21} N_{i2}^{*^2}
N^*_{j3} N^*_{j4} U^*_{k2} V^*_{k1} 
\nonumber \\
& - & 4 \delta M^n_{13} N^*_{i1} N^*_{i3} N^*_{j3} N^*_{j4}
U^*_{k1} V^*_{k1}-4 \delta M^n_{23} N^*_{i2} N^*_{i3}
N^*_{j3} N^*_{j4} U^*_{k1} V^*_{k1}-4 \delta M^n_{14}
N^*_{i1} N^*_{i4} N^*_{j3} N^*_{j4} U^*_{k1} V^*_{k1} 
\nonumber \\
& + &\delta m_{\tilde\chi^0_i} N_{j2}^{*^2} U^*_{k2} V^*_{k2}.
\eea

\section*{Appendix B}
\label{sec:appb}

Here we give $\delta M_1$, $\delta M_2$ and $\delta \mu$ for {\bf
  scheme 3}, obtained from solving equations (\ref{scheme3ab}).
Different variants of this scheme correspond to different choices of
$i,j,k$ (see Table~\ref{schemes}). The counterterms are given by
\bea
\delta M_{1} & = &\dfrac{N_{M_1}}{D},\\
\delta M_{2} & = &\dfrac{N_{M_2}}{D},\\
\delta \mu   & = & \dfrac{N_{\mu}}{2 D},
\eea
where,
\bea \label{D3}
D & = & (N^*_{i3} N^*_{i4} N^{*^2}_{j2} N^{*^2}_{k1}-N^{*^2}_{i2}
N^*_{j3} N^*_{j4} N^{*^2}_{k1}-N^*_{i3} N^*_{i4} N^{*^2}_{j1}
N^{*^2}_{k2}+N^{*^2}_{i1} N^*_{j3} N^*_{j4} N^{*^2}_{k2} 
\nonumber\\
& & +N^{*^2}_{i2} N^{*^2}_{j1} N^*_{k3} N^*_{k4}-N^{*^2}_{i1}
N^{*^2}_{j2} N^*_{k3} N^*_{k4}), 
\eea
\bea \label{dmu3}
N_{\mu} & = & - \delta m_{\tilde\chi^0_k} N^{*^2}_{i2} N^{*^2}_{j1} +  
\delta m_{\tilde\chi^0_k} N^{*^2}_{i1} N^{*^2}_{j2} +
   \delta m_{\tilde\chi^0_j} N^{*^2}_{i2} N^{*^2}_{k1} - 
  \delta m_{\tilde\chi^0_i} N^{*^2}_{j2} N^{*^2}_{k1}  
\nonumber\\
 & + & 2 \delta M^n_{13} N^*_{i1} N^*_{i3} N^{*^2}_{j2} N^{*^2}_{k1} + 
 2 \delta M^n_{23} N^*_{i2} N^*_{i3} N^{*^2}_{j2} N^{*^2}_{k1} + 
 2 \delta M^n_{14} N^*_{i1} N^*_{i4} N^{*^2}_{j2} N^{*^2}_{k1} 
\nonumber \\
 & + & 2 \delta M^n_{24} N^*_{i2} N^*_{i4} N^*_{j2} N^{*^2}_{k1} - 
 2 \delta M^n_{13} N^{*^2}_{i2} N^*_{j1} N^*_{j3} N^{*^2}_{k1} - 
 2 \delta M^n_{23} N^{*^2}_{i2} N^*_{j2} N^*_{j3} N^{*^2}_{k1} 
\nonumber \\ 
& - &  2 \delta M^n_{14} N^{*^2}_{i2} N^*_{j1} N^*_{j4} N^{*^2}_{k1} - 
 2 \delta M^n_{24} N^{*^2}_{i2} N^*_{j2} N^*_{j4} N^{*^2}_{k1} - 
  \delta m_{\tilde\chi^0_j} N^{*^2}_{i1} N^{*^2}_{k2} + 
  \delta m_{\tilde\chi^0_i} N^{*^2}_{j1} N^{*^2}_{k2} 
\nonumber \\ 
& - & 2 \delta M^n_{13} N^*_{i1} N^*_{i3} N^{*^2}_{j1} N^{*^2}_{k2} - 
 2 \delta M^n_{23} N^*_{i2} N^*_{i3} N^{*^2}_{j1} N^{*^2}_{k2} - 
 2 \delta M^n_{14} N^*_{i1} N^*_{i4} N^{*^2}_{j1} N^{*^2}_{k2} 
\nonumber \\ 
& - & 2 \delta M^n_{24} N^*_{i2} N^*_{i4} N^{*^2}_{j1} N^{*^2}_{k2} + 
 2 \delta M^n_{13} N^{*^2}_{i1} N^*_{j1} N^*_{j3} N^{*^2}_{k2} + 
 2 \delta M^n_{23} N^{*^2}_{i1} N^*_{j2} N^*_{j3} N^{*^2}_{k2} 
\nonumber \\ 
& + & 2 \delta M^n_{14} N^{*^2}_{i1} N^*_{j1} N^*_{j4} N^{*^2}_{k2} + 
 2 \delta M^n_{24} N^{*^2}_{i1} N^*_{j2} N^*_{j4} N^{*^2}_{k2} + 
 2 \delta M^n_{13} N^{*^2}_{i2} N^{*^2}_{j1} N^*_{k1} N^*_{k3} 
\nonumber \\ 
& - & 2 \delta M^n_{13} N^{*^2}_{i1} N^{*^2}_{j2} N^*_{k1} N^*_{k3} + 
 2 \delta M^n_{23} N^{*^2}_{i2} N^{*^2}_{j1} N^*_{k2} N^*_{k3} - 
 2 \delta M^n_{23} N^{*^2}_{i1} N^{*^2}_{j2} N^*_{k2} N^*_{k3} 
\nonumber \\ 
& + & 2 \delta M^n_{14} N^{*^2}_{i2} N^{*^2}_{j1} N^*_{k1} N^*_{k4} - 
 2 \delta M^n_{14} N^{*^2}_{i1} N^{*^2}_{j2} N^*_{k1} N^*_{k4} + 
 2 \delta M^n_{24} N^{*^2}_{i2} N^{*^2}_{j1} N^*_{k2} N^*_{k4} 
\nonumber \\ 
& - & 2 \delta M^n_{24} N^{*^2}_{i1} N^{*^2}_{j2} N^*_{k2} N^*_{k4},
\eea
\bea \label{dm23}
N_{M_2} & = & -\delta m_{\tilde\chi^0_k} N^*_{i3} N^*_{i4}
N^{*^2}_{j1}+\delta m_{\tilde\chi^0_k} N^{*^2}_{i1} N^*_{j3}
N^*_{j4}+\delta m_{\tilde\chi^0_j} N^*_{i3} N^*_{i4}
N^{*^2}_{k1} + \delta m_{\tilde\chi^0_i} N^{*^2}_{j1} N^*_{k3}
N^*_{k4} 
\nonumber \\ 
& - & 2 \delta M^n_{13}N^*_{i3} N^*_{i4} N^*_{j1} N^*_{j3}
N^{*^2}_{k1}-2 \delta M^n_{23}N^*_{i3} N^*_{i4} N^*_{j2}
N^*_{j3} N^{*^2}_{k1}-2 \delta M^n_{14}N^*_{i3} N^*_{i4}
N^*_{j1} N^*_{j4} N^{*^2}_{k1} 
\nonumber\\
& - & 2 \delta M^n_{24}N^*_{i3} N^*_{i4} N^*_{j2} N^*_{j4}
N^{*^2}_{k1}-\delta m_{\tilde\chi^0_i} N^*_{j3} N^*_{j4}
N^{*^2}_{k1}+2 \delta M^n_{13}N^*_{i1} N^*_{i3} N^*_{j3}
N^*_{j4} N^{*^2}_{k1} 
\nonumber \\ 
& + & 2 \delta M^n_{23}N^*_{i2} N^*_{i3} N^*_{j3} N^*_{j4}
N^{*^2}_{k1}+2 \delta M^n_{14}N^*_{i1} N^*_{i4} N^*_{j3}
N^*_{j4} N^{*^2}_{k1}+2 \delta M^n_{24}N^*_{i2} N^*_{i4}
N^*_{j3} N^*_{j4} N^{*^2}_{k1} 
\nonumber \\ 
& + & 2 \delta M^n_{13}N^*_{i3} N^*_{i4} N^{*^2}_{j1} N^*_{k1}
N^*_{k3}-2 \delta M^n_{13}N^{*^2}_{i1} N^*_{j3} N^*_{j4}
N^*_{k1} N^*_{k3}+2 \delta M^n_{23}N^*_{i3} N^*_{i4}
N^{*^2}_{j1} N^*_{k2} N^*_{k3} 
\nonumber \\ 
& - & 2 \delta M^n_{23}N^{*^2}_{i1} N^*_{j3} N^*_{j4} N^*_{k2}
N^*_{k3}+2 \delta M^n_{14}N^*_{i3} N^*_{i4} N^{*^2}_{j1}
N^*_{k1} N^*_{k4}-2 \delta M^n_{14}N^{*^2}_{i1} N^*_{j3}
N^*_{j4} N^*_{k1} N^*_{k4} 
\nonumber \\ 
& + & 2 \delta M^n_{24}N^*_{i3} N^*_{i4} N^{*^2}_{j1} N^*_{k2}
N^*_{k4}-2 \delta M^n_{24}N^{*^2}_{i1} N^*_{j3} N^*_{j4}
N^*_{k2} N^*_{k4}-\delta m_{\tilde\chi^0_j} N^{*^2}_{i1}
N^*_{k3} N^*_{k4}
\nonumber \\ 
& - &2 \delta M^n_{13}N^*_{i1} N^*_{i3} N^{*^2}_{j1} N^*_{k3}
N^*_{k4}-2 \delta M^n_{23}N^*_{i2} N^*_{i3} N^{*^2}_{j1}
N^*_{k3} N^*_{k4}- 2 \delta M^n_{14}N^*_{i1} N^*_{i4}
N^{*^2}_{j1} N^*_{k3} N^*_{k4}
\nonumber \\ 
& - & 2 \delta M^n_{24}N^*_{i2} N^*_{i4} N^{*^2}_{j1} N^*_{k3}
N^*_{k4}+2 \delta M^n_{13}N^{*^2}_{i1} N^*_{j1} N^*_{j3}
N^*_{k3} N^*_{k4} +  2 \delta M^n_{23}N^{*^2}_{i1} N^*_{j2}
N^*_{j3} N^*_{k3} N^*_{k4}
\nonumber\\ 
& + &2 \delta M^n_{14}N^{*^2}_{i1} N^*_{j1} N^*_{j4} N^*_{k3}
N^*_{k4}+2 \delta M^n_{24}N^{*^2}_{i1} N^*_{j2} N^*_{j4}
N^*_{k3} N^*_{k4}, 
\eea
and,
\bea \label{dm13}
N_{M_1} & = & \delta m_{\tilde\chi^0_k} N^*_{i3} N^*_{i4}
N^{*^2}_{j2}-\delta m_{\tilde\chi^0_k} N^{*^2}_{i2} N^*_{j3}
N^*_{j4}-\delta m_{\tilde\chi^0_j} N^*_{i3} N^*_{i4}
N^{*^2}_{k2}+2 \delta M^n_{13}N^*_{i3}  N^*_{i4} N^*_{j1}
N^*_{j3} N^{*^2}_{k2} 
\nonumber  \\
& + &2 \delta M^n_{23}N^*_{i3} N^*_{i4} N^*_{j2} N^*_{j3}
N^{*^2}_{k2}+2 \delta M^n_{14}N^*_{i3} N^*_{i4} N^*_{j1}
N^*_{j4} N^{*^2}_{k2}+2 \delta M^n_{24}N^*_{i3} N^*_{i4}
N^*_{j2} N^*_{j4} N^{*^2}_{k2} 
\nonumber \\
& + & \delta m_{\tilde\chi^0_i} N^*_{j3} N^*_{j4}
N^{*^2}_{k2}-2 \delta M^n_{13}N^*_{i1} N^*_{i3} N^*_{j3}
N^*_{j4} N^{*^2}_{k2}-2 \delta M^n_{23}N^*_{i2} N^*_{i3}
N^*_{j3} N^*_{j4} N^{*^2}_{k2} 
\nonumber \\
& - & 2 \delta M^n_{14}N^*_{i1} N^*_{i4} N^*_{j3} N^*_{j4}
N^{*^2}_{k2}-2 \delta M^n_{24}N^*_{i2} N^*_{i4} N^*_{j3}
N^*_{j4} N^{*^2}_{k2}-2 \delta M^n_{13}N^*_{i3} N^*_{i4}
N^{*^2}_{j2} N^*_{k1} N^*_{k3}
\nonumber \\ 
& + & 2 \delta M^n_{13}N^{*^2}_{i2} N^*_{j3} N^*_{j4} N^*_{k1}
N^*_{k3}-2 \delta M^n_{23}N^*_{i3} N^*_{i4} N^{*^2}_{j2}
N^*_{k2} N^*_{k3}+2 \delta M^n_{23}N^{*^2}_{i2} N^*_{j3}
N^*_{j4} N^*_{k2} N^*_{k3} 
\nonumber \\ 
& - & 2 \delta M^n_{14}N^*_{i3} N^*_{i4} N^{*^2}_{j2} N^*_{k1}
N^*_{k4}+2 \delta M^n_{14}N^{*^2}_{i2} N^*_{j3} N^*_{j4}
N^*_{k1} N^*_{k4}-2 \delta M^n_{24}N^*_{i3} N^*_{i4}
N^{*^2}_{j2} N^*_{k2} N^*_{k4} 
\nonumber \\
& + &2 \delta M^n_{24}N^{*^2}_{i2} N^*_{j3} N^*_{j4} N^*_{k2}
N^*_{k4}+\delta m_{\tilde\chi^0_j} N^{*^2}_{i2} N^*_{k3}
N^*_{k4}-\delta m_{\tilde\chi^0_i} N^{*^2}_{j2} N^*_{k3}
N^*_{k4} 
\nonumber \\ 
& + & 2 \delta M^n_{13}N^*_{i1} N^*_{i3} N^{*^2}_{j2} N^*_{k3}
N^*_{k4}+2 \delta M^n_{23}N^*_{i2} N^*_{i3} N^{*^2}_{j2}
N^*_{k3} N^*_{k4}+2 \delta M^n_{14}N^*_{i1} N^*_{i4}
N^{*^2}_{j2} N^*_{k3} N^*_{k4} 
\nonumber \\
& + & 2 \delta M^n_{24}N^*_{i2} N^*_{i4} N^{*^2}_{j2} N^*_{k3}
N^*_{k4}-2 \delta M^n_{13}N^{*^2}_{i2} N^*_{j1} N^*_{j3}
N^*_{k3} N^*_{k4}-2 \delta M^n_{23}N^{*^2}_{i2} N^*_{j2}
N^*_{j3} N^*_{k3} N^*_{k4} 
\nonumber \\ 
& - & 2 \delta M^n_{14}N^{*^2}_{i2} N^*_{j1} N^*_{j4} N^*_{k3}
N^*_{k4}-2 \delta M^n_{24}N^{*^2}_{i2} N^*_{j2} N^*_{j4}
N^*_{k3} N^*_{k4}. 
\eea

\section*{Appendix C}
\label{sec:appc}

We show by explicit calculation that using the $\tilde\chi_i^+
\tilde\chi_i^- Z$ couplings, it is possible to determine which of the two
charginos is  more wino--like, or, equivalently, whether $M_2$ is
bigger or smaller than $|\mu|$. The relevant terms we compare are
\beq \label{comp1}
C_{i} = (-V_{i1}^2-\dfrac{1}{2} V_{i2}^2+ s_W^2)^2 + 
(-U_{i1}^2 - \dfrac{1}{2} U_{i2}^2+ s_W^2)^2, ~\forall~ i \in
~\{1,2\} \, ,
\eeq
which determine the sum of the squares of the left-- and right--handed
couplings.
These terms contribute, for example, to the chargino pair production
cross section through $Z-$boson exchange in the $s-$channel in $e^{+}
e^{-}$ or $pp$ collisions. Recall that we use the convention $M_2 > 0$
and assume vanishing CP phases in the chargino mass matrix. However,
$\mu$ can be either positive or negative. The unitary mixing matrices
$U$ and $V$ are real. They satisfy
\bea \label{ucond}
U_{ii}^{2} = U_{jj}^{2},~ U_{ij}^{2} = U_{ji}^{2},\nonumber\\
V_{ii}^{2} = V_{jj}^{2},~ V_{ij}^{2} = V_{ji}^{2},
\eea
where $\{i,j\} \in \{1,2\}$. 

Recall that, for given $\tan\beta$, we can solve the equations for the
physical masses $m_{\tilde \chi_i^+}$ for the parameters $M_2$ and
$\mu$, with a four--fold ambiguity. Part of this ambiguity is due to
the fact that the expressions for the physical masses are invariant
under the exchange of $M_2$ and $\mu$. Here we show that the
quantities $C_i$ of Eq.(\ref{comp1}) can be used to resolve this
ambiguity. We show this separately for positive and negative $\mu$.

\begin{itemize}
\item $case ~1 :$  $\mu > 0$\\
  Let $a, b$ be the possible solutions for $M_2, \mu$; $a,b > 0$.  For
  $\{M_2, \mu\} = \{a,b\}$, let $\mathcal{U}$ and $\mathcal{V}$ be the
  unitary mixing matrices satisfying Eq.(\ref{mcd}). Then, for
  $\{M_2, \mu\} = \{b,a\}$, the corresponding matrices are given by
  $\mathcal{VO}$ and $\mathcal{UO}$ respectively, where
\beq \label{trans1}
  \cal{O} = \left( \begin{array}{cc}
      0 & 1 \\
      1 & 0
\end{array} 
\right).
\eeq
Using relations (\ref{ucond}), it is straightforward to observe that 
in $M^c$, $a \leftrightarrow b$ corresponds to $C_{1} \leftrightarrow
C_{2}$.

\item $case ~2 :$  $\mu < 0$\\ 
Let $ a, b$ be the possible solutions for $M_2, |\mu|$; $a,b > 0$. 
For $\{M_2, \mu\} = \{a,-b\}$, let $\mathcal{U}$ and $\mathcal{V}$ be the 
unitary mixing matrices satisfying Eq.(\ref{mcd}). Then, for
$\{M_2, \mu\} = \{b,-a\}$, the corresponding matrices are given by
$\mathcal{V\tilde{O}}$ and $\mathcal{U\tilde{O}}^{T}$ respectively, where
\beq \label{trans2}
\cal{\tilde{O}} =  \left( \begin{array}{cc}
0 & 1 \\
-1 & 0 
\end{array} 
\right).
\eeq
Again, $\{a,-b\} \leftrightarrow \{b,-a\}$ corresponds to $C_{1}
\leftrightarrow C_{2}$.
\end{itemize} 

We thus see that by comparing $C_{1}$ with $C_{2}$ it is possible in
both cases to determine the ordering of $M_2$ and $|\mu|$. However, if $M_2
\simeq |\mu|$, it can be difficult to infer whether $M_{2} >
|\mu|$ or vice versa. This is of some importance, since precisely in
this region of parameter space schemes where the mass of the more
higgsino--like chargino is used as input show an instability, as we
saw in Sec.~4.

\bibliographystyle{utphys}
\addcontentsline{toc}{section}{References}
\bibliography{ourbib}
\end{document}